\documentclass[aip,jcp,reprint,groupedaddress,floatfix]{revtex4-1}

\usepackage{hyperref}
\usepackage{subeqnarray}
\usepackage{comment}
\usepackage{amsmath,amssymb,bbm}
\usepackage{graphics,graphicx}
\usepackage{mathrsfs}
\usepackage{dsfont}
\usepackage{cleveref}
\usepackage[dvipsnames]{xcolor}
\usepackage{epstopdf}
\usepackage[english]{babel}

\newcommand*{\ie} {i.e.}
\newcommand*{\eg} {e.g.}
\newcommand*{\vs} {vs.}

\newcommand*{\trm}{\textrm}

\newcommand*{\eq} [1]{Eq.~(\ref{#1})}
\newcommand*{\eqs}[1]{Eqs.~(\ref{#1})}

\newcommand*{\reff}[1]{(\ref{#1})}
\newcommand*{\citen}[1]{Ref.~[\!\!\citenum{#1}]}

\renewcommand*{\sec}[1]{Section~\ref{#1}}

\newcommand{\Dop}{\hat{D}}
\newcommand{\nocc}{\rho}
\newcommand{\Ne} {N}
\newcommand{\Fop}     {\hat{H}}
\newcommand{\efermi}{\epsilon_{\trm{f}}}
\newcommand{\kb} {k_\trm{B}}
\newcommand{\trace} [1]{\trm{Tr}\{#1\}}
\newcommand{\Dm}  {D} 
\newcommand{\dm}  {P}
\newcommand{\Nb} {\bar{N}}
\newcommand{\Db}    {\bar{\Dm}}
\newcommand{\db}{\bar{P}}
\newcommand{\Id }{I} 
\newcommand{\Dmp} {\Dm_{\pert}}
\newcommand{\pert}{\lambda}
\newcommand{\Fmp} {\Fm_{\pert}}
\newcommand{\real}{\mathbb{R}}
\newcommand{\Fm}  {H}
\newcommand{\Dmk} [1]{\Dm^{(#1)}} 
\newcommand{\Pmp} {H}
\newcommand{\Fmk}[1]{\Pmp^{(#1)}}
\newcommand{\Pmk}  [3]{#1_{#2}^{(#3)}} 
\newcommand{\op }   {\hat{O}}        
\newcommand{\mop }{O}        

\newcommand{\Dmovk} [1]{\Dm_{ov}^{(#1)}}
\newcommand{\Dmovijk} [1]{\Dm_{ov,ij}^{(#1)}}
\newcommand{\Dmvok} [1]{\Dm_{vo}^{(#1)}}
\newcommand{\Dmvvk} [1]{\Dm_{vv}^{(#1)}}
\newcommand{\Dmook} [1]{\Dm_{oo}^{(#1)}}

\newcommand{\Fmovk}   [1]{\Pmp_{ov}^{(#1)}}
\newcommand{\Fmovijk} [1]{\Pmp_{ov,ij}^{(#1)}}
\newcommand{\Fmvok} [1]{\Pmp_{vo}^{(#1)}}

\newcommand{\Em} {\epsilon}
\newcommand{\Eb} {\bar{\Em}}

\newcommand{\Kpoly}[1]{\mathscr{F}_{#1}}        
\newcommand{\Ipoly} [1]{\mathscr{L}_{#1}}        
\newcommand{\etmax} {\tilde{\epsilon}_\trm{max}}
\newcommand{\etmin}  {\tilde{\epsilon}_\trm{min}}
\newcommand{\emax} {\epsilon_\trm{max}}
\newcommand{\emin}  {\epsilon_\trm{min}}
\newcommand{\sign}     {\trm{sign}}
\newcommand{\xflex}{x_\trm{flex}}
\newcommand{\N}  {M}
\newcommand{\Kpolypert}[1]{\mathscr{F}^{(k)}_{#1}}        
\newcommand{\Ipolypertb} [1]{\mathscr{L}^{(k>0)}_{#1}}        
\newcommand{\Kpolypertk}[2]{\mathscr{F}^{(#1)}_{#2}}        
\newcommand{\Ipolypertk} [2]{\mathscr{L}^{(#1)}_{#2}}        
\newcommand{\Dbk}[1]{\bar{\Dm}^{(#1)}}

\begin{document}

\title{Notes on density matrix perturbation theory}


\author{Lionel A. Truflandier}
\email{lionel.truflandier@u-bordeaux.fr}
\affiliation{Institut des Sciences Mol\'{e}culaires (ISM), Universit\'{e} Bordeaux,
CNRS UMR 5255, 351 cours de la Lib\'{e}ration, 33405 Talence cedex, France}

\author{Rivo M. Dianzinga}
\affiliation{Institut des Sciences Mol\'{e}culaires (ISM), Universit\'{e} Bordeaux,
CNRS UMR 5255, 351 cours de la Lib\'{e}ration, 33405 Talence cedex, France}

\author{David R. Bowler}
\affiliation{London Centre for Nanotechnology, UCL, 17-19 Gordon St, 
London WC1H 0AH, Department of Physics \& Astronomy, UCL, Gower St, 
London, WC1E 6BT, UK}
\affiliation{International Centre for Materials Nanoarchitechtonics (MANA), 
National Institute for Materials Science (NIMS), 1-1 
Namiki,Tsukuba, Ibaraki 305-0044, Japan}

\date{\today}


\begin{abstract}
Density matrix perturbation theory (DMPT) is known as a promising alternative
to the Rayleigh-Schr{\"o}dinger perturbation theory, in which the sum-over-state (SOS) is
replaced by algorithms with perturbed density matrices as the input variables. In this
article, we formulate and discuss three types of DMPT, with two of them based only
on density matrices: the approach of Kussmann and Ochsenfeld [J. Chem. Phys.127,
054103 (2007)] is reformulated via the Sylvester equation, and the recursive DMPT of
A.M.N. Niklasson and M. Challacombe [Phys. Rev. Lett. 92, 193001 (2004)] is extended
to the hole-particle canonical purification (HPCP) from [J. Chem. Phys. 144, 091102 (2016)]. 
Comparison of the computational performances shows that the aforementioned methods 
outperform the standard SOS. The HPCP-DMPT demonstrates stable convergence profiles 
but at a higher computational cost when compared to the original recursive polynomial method.
\end{abstract}

\maketitle

\section{Introduction}

Traditionally, analytical evaluation of the response of a system to a perturbation is based on the Rayleigh-Schr{\"o}dinger 
perturbation theory (RSPT), taking the form of a sum-over-states (SOS) which requires knowledge of the full set of eigenstates. 
Recent emerging developments have seen the SOS replaced with the response equation, resolved using density matrices 
as the working variables along with matrix-matrix multiplication-rich recursion algorithms such as kernel 
polynomials.\cite{kernelpol_2006}
Working directly with density matrices is of great interest since we can exploit their natural property 
of sparsity which is the keypoint in designing linear scaling approaches (though we leave the application of sparsity to this work to a future publication). Another advantage is found in the fact that matrix-matrix multiplication can be efficiently 
parallelized using MPI or optimized on GPUs.

The first applications of the RSPT to molecular-orbital (MO) wavefunction-based 
self-consistent-field (SCF) methods were introduced during the 60s for the computation
of molecular properties such as magnetic susceptibility,\cite{stevens_jcp_1963}
static polarisabilities and force constants,\cite{gerrattmills_jcp_1968-1,gerrattmills_jcp_1968-2}
which are all related to second-order energy derivatives through the calculation of the
first-order change of the wavefunctions with respect to the perturbation.
Similarly to the unperturbed case, variational solutions of the perturbed MOs are obtained by solving the 
so-called coupled-perturbed self-consistent field (CPSCF) 
equations.\cite{thomsen_molphys_1973,ditchfield_molphys_1974,pople_ijqc_1979} 
These early developments based either on the perturbed molecular-orbitals or mixed perturbed 
atomic-orbitals/molecular-orbitals (AOs/MOs) were well-known to involve cumbersome ---especially at that time--- 
matrix transformations.\citep{frisch_cp_1990,osamura_theochem_1983}
In 1962, McWeeny had already introduced the elegant formalism of the
density matrix perturbation theory\cite{mcweeny_physrev_1962,mcweeny_cpl_1968} 
(DMPT) ---extended to the CPSCF equations resolution by 
Diercksen and McWeeny\cite{diercksen_jcp_1966} for the evaluation of $\pi$-electron
polarizabilities using the Pariser-Parr-Pople model.\cite{ppp_pople_1953,ppp_pariser1_1953,ppp_pariser2_1953}
Note that the DMPT formulation of McWeeny still required a SOS and thus the knowledge of the eigenstates.
This work has first inspired Moccia to generalize the McWeeny-CPSCF equations resolution 
to non-orthogonal basis.\cite{moccia_tca_1967,moccia_cpl_1970-1}
Perturbation-dependent non-orthogonal basis implementation was then proposed by Dodds, McWeeny, Sadlej 
and Wolinski\cite{dodds_molphys_1977-1,dodds_molphys_1977-2,wolinski_molphys_1980} for the calculation of 
atomic (hyper)-polarisabilities using the Hartree-Fock method in conjunction with Gaussian-type orbital basis sets. 
The advantages of the McWeeny's approach over AOs/MOs-CPSCF have been clearly outlined, for instance, 
in the seminal article of Wolinski, Hinton, and Pulay\cite{wolinski_jacs_1990} dealing with the calculation of magnetic shieldings.

In comparaison to the RSPT-SOS, density matrix based methods were introduced more recently, around the years 2000.
Ochsenfeld and Head-Gordon first reformulate the CPSCF equations in terms of the density 
matrix only\cite{ochsenfeld_cpl_1997} (referred to as D-CPSCF by the authors) starting from the 
Li-Nunes-Vanderbilt (LNV) unconstrained energy functional\cite{LNV_PRB_1993} where the 
McWeeny purification polynomial\cite{McWeeny_density_1956a,McWeeny_RevModPhys_1960} 
is used as input density matrix. Later, Kussmann and Ocshenfeld recognized important deficiencies in this 
initial version which was corrected for in the alternative derivation of \citen{kussmann_jcp_2007-1,kussmann_jcp_2007-2}. 
Within the same spirit, Lasen et al.\cite{larsen_jcp_2001-1} followed by Coriani et al.\cite{coriani_jcp_2007}
have derived and implemented, respectively, response equations using the asymmetric 
Baker-Campbell-Hausdorff expansion\cite{helgaker_cpl_2000} for the auxiliary density matrix.
Within the field of density matrix purifications, Niklasson, Weber and Challacombe have introduced
a recursive variant of the DMPT\cite{niklasson_prl_2004,niklasson_jcp_2005,niklasson_jcp_2007} 
based on the purification spectral projection method detailed in \citen{Niklasson_TCP_2002}. 
Whereas their theoretical framework is general, that is, any recursive polynomial expansion respecting 
constraints imposed by the density matrix properties may be considered, only performances of the second-order 
polynomial trace-correcting purification\cite{Niklasson_TCP_2002} (TC2)
---also referred to as second-order spectral projection\cite{Cawkwell_JCTC_2012,Cawkwell_JCTC_2014}---  
have been investigated.\cite{weber_prl_2004,weber_jcp_2005}

Recently, we have introduced a Lagrangian formulation for the constrained minimization of the 
$N$-representable density matrix\cite{truflandier_jcp_2016} based on the McWeeny idempotency 
error functional.\cite{McWeeny_density_1956a} Within the canonical ensemble (\textit{NVT}),
this gave rise to a unique trace-conserving recursive polynomial purification which can be recast
in terms of the the hole-particle duality condition. The closed-form of this hole-particle canonical 
purification (HPCP) makes it self-consistent, that is, heuristic adjustment of the polynomial during 
the course of the purification is not required. Moreover, providing an adequate initial guess,
the HPCP is variational and monotonically convergent.\cite{truflandier_jcp_2016}

Following the pioneer work of Niklasson and Challacombe\cite{niklasson_prl_2004}
our current aim is to develop a robust and performant purification based DMPT using
the HPCP. In this paper we are mainly concerned to ($i$) review the SOS-McWeeny-DMPT,
the Kussmann and Ocshenfeld formulation of DMPT (later referred to as Sylvester-DMPT,
\textit{vide infra}) and the purification-DMPT, ($ii$) derive DMPT equations based on the 
HPCP density matrix kernel using an orthogonal representation, and ($iii$) perform a fair 
comparison of the computational efficiency of the aforementioned methods.

\section{Theoretical background}
\label{sec:theorback}

\subsection{The one-electron density matrix}
\label{sec:dm}

We consider an ensemble of fermions at the themodynamical equilibrium in the external potential 
created by the nuclei. Given a set of $\Ne$ occupied states, whose wave-function is written in the 
form of a single determinant, the general expression for the spinless one-particle density operator is
\begin{equation}
   \Dop = \sum_i \nocc_i |\psi_i\rangle\langle\psi_i|
   \label{eq:densop}
\end{equation}
where $\{\nocc_i\}$ are occupation numbers associated with the one-electron states 
$\{\psi_i | \langle\psi_{i}|\psi_{j}\rangle=\delta_{ij}\}$, the latter being for instance eigenvectors 
of any one-electron model Hamiltonian in tight-binding approach or Fock operator when dealing
with a self-consistent field method as found in the Hartree-Fock or Kohn-Sham mean-field approximation.
Thereafter, it shall be denoted by $\Fop$, irrespective of the approach. Given $\Fop$, for $\Dop$ to 
describe a stationary state within the \textit{NVT} ensemble, the necessary and sufficient conditions are:
\begin{subequations}
\label{eq:nrep}
 \begin{align}
    [\Fop,\Dop] = 0,\quad\trm{subject to}\label{eq:stationary}\\
    \Dop          = \Dop^{\dagger} \label{eq:hermi} \\
    0\leq   \nocc_i \leq 1             \label{eq:idemp} \\
	\sum_i\nocc_i = \Ne                   \label{eq:trace}   
 \end{align}
\end{subequations}
where $[\cdot,\cdot]$ is the usual symbol for the commutator of two operators. Note that the hermicity 
constraint in (\ref{eq:hermi}) is already enforced by the definition (\ref{eq:densop}). But, if we are interested 
in solving $\Dop$ directly, without the support of the eigenstates, this condition will have to be imposed at 
the beginning and during the iterative resolution. As a result, \eq{eq:nrep} mainly expresses that $\Fop$ and 
$\Dop$ must share the same eigenstates subject to the $\Ne$-representability constraints, which are:  ($i$) 
no more than two electron can occupy a given state, assuming spin paired electrons, ($ii$) the total number 
of electrons ($2\Ne$) is fixed. If now, we want to guarantee that $\hat{D}$ corresponds to 
the ground state, \ie\ the lowest energy states are filled up to the Fermi level ($\efermi$) and allows for
fractional occupation around $\efermi$, we should combine conditions (\ref{eq:nrep}) with 
the Fermi-Dirac (FD) distribution, 
\begin{equation}
	 \Dop = \left( \Id + e^{\beta(\Fop -\mu\Id)} \right)^{-1} 
	 \label{eq:fermidirac} 
\end{equation}
where $\beta=(\kb T)^{-1}$ is the inverse electron temperature and the chemical  potential, $\mu$, 
is chosen to conserve the number of electrons. Note that definition \reff{eq:fermidirac} is a substitute 
to \eq{eq:densop} if we want to circumvent explicit calculation of the eigenstates. The FD distribution 
also demonstrates that, for non-degenerate systems, there exists a correspondence between the density 
and the Hamiltonian operator. Unfortunatly its non-linear character prevents a direct resolution of $\Dop$ in 
terms of $\Fop$. This has motivated the introduction of the density matrix polynomial expansion in the 
90's to resolve $\Dop$ recursively with $\Fop$ as input.\cite{Goedecker_PRB_1995,Baer_PRL_1997}
In this work, we shall reduce the theoretical framework to pure states 
where the occupation numbers of the eigenstates are either 0 or 1. This leads to
\begin{equation}
   \Dop = \sum_{i\in\trm{occ}} |\psi_i\rangle\langle\psi_i| 
   \label{eq:densop_idemp}
\end{equation}
where, compared to \eq{eq:densop}, the subset of occupied states is sufficient to fully determine 
the one-particle density matrix. In that case, the $\Ne$-representability conditions of 
\eqs{eq:idemp}--\reff{eq:trace} can be recast as  
\begin{subequations}
\label{eq:nrep_idemp}
\begin{align}
     \Dop            &= \Dop^2\label{eq:nrep_idempa}\\
     \trace{\Dop}&= \Ne      \label{eq:nrep_idempb}
\end{align}
\end{subequations}
that is, the density matrix of pure state is idempotent. The corresponding ground-state 
is determined by the zero temperature limit of \eq{eq:fermidirac}, 
\begin{equation}
	\Dop = \Theta(\mu I - \Fop)\quad\trm{with}\quad\lim_{T\rightarrow 0}{\mu} = \efermi
	\label{eq:fermidirac_idemp} 
\end{equation}
with $\Theta$ the Heaviside step function. If now we consider a separable Hilbert space 
of dimension $M$ $(\geq \Ne)$ which admits an orthonormal basis 
$\{\phi_{\mu}\in L^2 | \langle\phi_{\mu}|\phi_{\nu}\rangle=\delta_{\mu\nu}\}_{\mu=1}^{M}$, 
the one-particle density operator of \eq{eq:densop_idemp} has the following matrix representation
\begin{equation}
	\Dm = \sum_{i\in\trm{occ}}\dm_i\quad\trm{with}\quad\dm_i = c^{ }_{i}\otimes c_{i}
	\label{eq:densmat_elecd}
\end{equation}
where $\{c_i\in\real^{M\times 1}|c^{\dagger}_{i}c^{ }_{j}=\delta_{ij}\}_{i=1}^{\Ne}$ 
are the column vectors containing the expansion coefficients such that,
$\{|\psi_i\rangle = \sum^M_{\mu} c_{\mu i}|\phi_{\mu}\rangle| c_{\mu i}
=\langle\phi_{\mu}|\psi_i\rangle\}_{i=1}^{N}$, and $\{\dm_i\in\real^{M\times M}
 |\dm_{i}\dm_{j}=\dm_{i}\delta_{ij},\trace{\dm_i}=1\}_{i=1}^{N}$ 
is the set of the $\Ne$ orthonormal projectors belonging to the subset of
occupied states. At this stage, we shall introduce the one-hole density matrix 
built from the subspace of the $\Nb$ virtual (unoccupied) states 
\begin{equation}
	\Db = \sum_{i\in\trm{virt}}\db_i\quad\trm{with}\quad\db_i = \bar{c}^{ }_{i}\otimes\bar{c}_{i}
	\label{eq:densmat_hole}
\end{equation}
such that $M=\Nb+\Ne$. Throughout the paper, quantities related to those states will be indicated 
by a bar accent. Stationary one-particle and one-hole density matrices must obey the two following identities: 
\begin{eqnarray} 
	&\Dm  + \Db = \Id  \label{eq:closure}\\
	&\Dm\Db      = 0    \label{eq:orthog}
\end{eqnarray} 
with $\Id$ the identity matrix. As a result, it can be easily demonstrated that the one-hole 
density matrix for a pure state obeys the same properties as its one-particle equivalent, 
\eg\ the idempotency and trace conservation of \eqs{eq:nrep_idemp}.
\subsection{One-electron density matrix perturbation theory}
\label{sec:dmpt}
Let us now consider the perturbed one-particle density and Hamiltonian matrix $\Dmp$ and $\Fmp$, 
respectively, where $\lambda$ stands for any time-independent perturbation. 
At the zero electronic temperature limit, for $\Dmp$ to describe the perturbed stationary 
state corresponding to the unperturbed ground state $\Dm$, it must also obey the
following rules:
\begin{subequations}
\label{eq:nrep_pert}
 \begin{align}
    [\Fmp,\Dmp] = 0,\quad\trm{subject to:}\label{eq:stationary_pert}\\
    \Dmp     = \Dmp^{\dagger}      \label{eq:hermi_pert} \\
    \Dmp     = \Dmp^2      \label{eq:idemp_pert} \\
    \trace{\Dmp} = N    \label{eq:trace_pert}       
 \end{align}
\end{subequations}
where \eqs{eq:idemp_pert} and \reff{eq:trace_pert} stands for the \textit{N}-representability conditions.
Note that \reff{eq:hermi_pert} will be enforced by construction (\textit{vide infra}). We shall expand the perturbed 
density and Hamiltonian matrix in power series with respect to a perturbation parameter ($0<\pert\leq 1$),
\begin{subequations} 
\begin{align}
  \Dmp &= \Dmk{0} + \lambda\Dmk{1} + \lambda^2\Dmk{2} + \ldots + \lambda^k\Dmk{k}   \label{eq:Dmpert} \\
  \Fmp &= \Fmk{0} + \lambda\Fmk{1} + \lambda^2\Fmk{2} + \ldots  + \lambda^k\Fmk{k}   \label{eq:Fmpert} 
\end{align}
\end{subequations}
where $\Pmk{X}{}{k}$ represents the $k$th-order change of the quantity $X$ 
with respect to $\lambda$; $\Dmk{0}\equiv\Dm$ and $\Fmk{0}\equiv\Fm$ are the unperturbed
density and Hamiltonian matrix, respectively. On inserting the expansion \reff{eq:Dmpert} 
into the $N$-representability constraints of \eqs{eq:idemp_pert} and 
\reff{eq:trace_pert}, and by equating the perturbation orders, 
this yields to:
\begin{subequations} 
\label{eq:idemexp}
\begin{align} 
  &\Dm^{2} = \Dm\nonumber\\ 
  &\quad\trm{subject to}\quad\trace{\Dm}=N, \label{eq:idemexp0}\\
  &\Dm\Dmk{1} + \Dmk{1}\Dm = \Dmk{1}\nonumber\\
  &\quad\trm{subject to}\quad\trace{\Dmk{1}}=0,  \label{eq:idemexpa}\\  
  &\Dm\Dmk{2} + (\Dmk{1})^2 + \Dmk{2}\Dm = \Dmk{2},\nonumber\\
  &\quad\trm{subject to}\quad\trace{\Dmk{2}}=0, \label{eq:idemexpb}\\
  &\Dm\Dmk{3} + \Dmk{1}\Dmk{2} + \Dmk{2}\Dmk{1} + \Dmk{3}\Dm = \Dmk{3}\nonumber\\
  &\quad\trm{subject to}\quad\trace{\Dmk{3}}=0,  \label{eq:idemexpc}\\ 
  &\quad\quad\quad\vdots \nonumber\\
  &\sum^{k}_{l=0} \Dmk{l}\Dmk{k-l} =\Dmk{k}\nonumber\\
  &\quad\trm{subject to}\quad\trace{\Dmk{k}}=0\label{eq:idemexpg}
\end{align}
\end{subequations} 
Further repeating the perturbation identification by introducing \eqs{eq:Dmpert} and \reff{eq:Fmpert} in
the commutator of \eq{eq:stationary_pert}, we obtain:
\begin{subequations} 
\label{eq:scfexp}
\begin{align}
   &[\Fm,\Dm] = 0 \label{eq:scfexp0} \\
   &[\Fm,\Dmk{1}] + [\Fmk{1},\Dm] = 0   \label{eq:scfexpa} \\ 
   &[\Fm,\Dmk{2}] + [\Fmk{1},\Dmk{1}] + [\Fmk{2},\Dm] = 0  \label{eq:scfexpb} \\
   &[\Fm,\Dmk{3}] + [\Fmk{1},\Dmk{2}] + [\Fmk{2},\Dmk{1}] + [\Fmk{3},\Dm] = 0  \label{eq:scfexpc} \\ 
   &\quad\quad\quad\vdots \nonumber\\
   &\sum^{k}_{l=0} [\Fmk{l},\Dmk{k-l}] = 0 \label{eq:scfexpg} 
\end{align}
\end{subequations} 
The generalized idempotency constraint of \eq{eq:idemexpg} and the stationary condition 
of \eq{eq:scfexpg} constitute the working equations for developing the various forms of
the density matrix perturbation theory (DMPT) which are described in next sections.

Derivation of the expressions for the observable quantities induced by 
perturbations as energy contributions in the power series expansion
$E_{\lambda} = E^{(0)} + \lambda E^{(1)} + \lambda^2 E^{(2)} + \ldots  + \lambda^k E^{(k)}$,
can be found in many text books and references.\cite{hirschfelder_recent_1964,helgaker_molecular_2000}
In the case of the tight-binding method, the unperturbed one-electron 
energy ($E^{(0)}\equiv E$) is simply
\begin{equation} 
E = 2\trace{HD}
\end{equation}
By stopping at the first-order in the Hamiltonian expansion of \eq{eq:Fmpert}, it can be 
shown that the corresponding energy corrections for $k > 0$ are given by\cite{niklasson_prl_2004} 
\begin{equation} 
E^{(k+1)} = \frac{2}{k}\trace{H^{(1)}D^{(k)}}
\label{eq:fockener_pert}
\end{equation}
Consequently, energy derivatives up to the order $(k+1)$ involves 
knowledge of the density matrices up to the order $k$.
A more popular approach,\cite{pulay_second_1983,
sekino_frequency_1986,karna_frequency_1991,furche_density_2001,
kobayashi_dynamic_2012,kussmann_jcp_2007-2} 
for computing energy derivatives for $k>2$, relies on the $(2k+1)$ Wigner rule,  
which states that $E^{(2k+1)}$ can be obtained from the $k$th-order perturbed 
wavefunctions.\cite{hirschfelder_recent_1964} It is noteworthy that McWeeny 
et al.\cite{mcweeny_physrev_1962,diercksen_jcp_1966,dodds_molphys_1977-1,
dodds_molphys_1977-2} ---later reported by Niklasson et al.\cite{niklasson_prl_2004,
niklasson_jcp_2005,niklasson_jcp_2007}--- have adapted the $(2k + 1)$th thereom 
to perturbed density matrices as inputs, up to order 4 with respect to the energy, 
without any support of the wavefunctions.
\subsection{SOS-McWeeny-DMPT}
\label{sec:ao-wcpscf}
The DMPT equations proposed by McWeeny involve the partioning of 
$\Dmk{k}$ into four distinct contributions and their resolutions.\cite{mcweeny_physrev_1962}  
Using the closure relation of \eq{eq:closure}, any matrix representation of the operator $\op$
can be expressed into the following projected components:
\begin{eqnarray}
    &\mop = \mop_{oo} + \mop_{ov} + \mop_{vo} + \mop_{vv}  \label{eq:mcwresolve}\\
	\trm{with:}&\nonumber\\
	&\mop_{oo}=\Dm\mop\Dm\nonumber\\
	&\mop_{vv}=\Db\mop\Db\nonumber\\
	&\mop_{ov}=\Dm\mop\Db\nonumber\\
	&\mop_{vo}=\Db\mop\Dm\nonumber
\end{eqnarray}
The subscripts $oo$ and $vv$ designate the occupied-occupied and virtual-virtual 
diagonal contributions whereas $ov$ and $vo$ stand for the non-diagonal occupied-virtual
and virtual-occupied transition terms. To the first order of perturbation, 
on applying the projection decomposition of \eq{eq:mcwresolve} to both sides of
\eq{eq:idemexpa},\cite{mcweeny_physrev_1962} we obtain:
\begin{equation}
	2\Dmook{1} + \Dmovk{1} + \Dmvok{1} = \Dmook{1} + \Dmovk{1} + \Dmvok{1} + \Dmvvk{1}\label{eq:mcwresolved}
\end{equation}
where, by comparing terms of the left-hand and right-hand sides (abbreviated by 
lhs and rhs, respectively, in the rest of the text), it can be easily deduced that
\begin{equation}
	\Dmook{1} = 0,\quad\Dmvvk{1} = 0  \label{eq:mcwresolvee}
\end{equation}
As a result, the first-order perturbed density matrix is fully determined by the 
occupied-virtual transition matrix, such that
\begin{equation}
	\Dmk{1} = \Dmovk{1} + \Dmvok{1} = \Dmovk{1} + (\Dmovk{1})^{\dagger}  \label{eq:mcwresolveg}
\end{equation}
Resolving $\Fmk{1}$ into the four components using \eq{eq:mcwresolve}, 
we can search for $\Dmovk{1}$ through \eq{eq:scfexpa}. After simplification, 
the following working equations are found	
\begin{equation}
\label{eq:mcwfstorder}
  \Fmovk{1} = [\Fm,\Dmovk{1}]\quad\trm{and}\quad\Fmvok{1} = [\Dmvok{1},\Fm] 
\end{equation}   
On recalling the Hermitian property of the unperturbed and perturbed 
Hamiltonian matrices, the lhs of \eq{eq:mcwfstorder} is found to be the conjugate 
transpose of the rhs, then solving one of the two equation is sufficient 
to evaluate the perturbed density matrix of \eq{eq:mcwresolveg}. 

The common Rayleigh-Schr{\"o}dinger sum-over-states (SOS) is recovered from 
\eq{eq:mcwfstorder}, by applying the spectral resolution for the 
non-perturbed Hamiltonian matrix according to
\begin{equation} 
	\Fm = \sum^{}_{i\in\trm{occ}} \Em_i \dm_i + \sum^{}_{j\in\trm{virt}} \Eb_j \db_j 
	\label{eq:mcweenyfocka}
\end{equation}
where indices $i$ and $j$ run over the energy-weighted projectors for the occupied
and unoccupied space, respectively. On substitution of \eqs{eq:mcweenyfocka} 
into \eq{eq:mcwfstorder}, using the following identity,
\begin{equation} 
	\mop_{ov}=\sum_{i\in\trm{occ}}\sum_{j\in\trm{virt}}\dm_i\mop\db_j
	\label{eq:ov_identity}
\end{equation}
we obtain:
\begin{subequations}
\begin{align}
  \sum^{}_{i\in\trm{occ}}\sum^{}_{j\in\trm{virt}} &\left(\Dmovijk{1} (\Em_i - \Eb_j) - \Fmovijk{1} \right) = 0\\
  \trm{with:}&\ \ \ \Dmovijk{1} := \left(\dm_i\Dmk{1}\db_j\right)\in\real^{M\times M} \\
                  &\ \ \ \Fmovijk{1}  := \left(\dm_i\Fmk{1}\db_j\right)\in\real^{M\times M}    
\end{align}
\end{subequations}
This equation can be recast into the following SOS form:
\begin{equation}
	\Dmovk{1} = \sum^{}_{i\in\trm{occ}}\sum^{}_{j\in\trm{virt}}\frac{\Fmovijk{1}}{\Em_i - \Eb_j} 
	\label{eq:sosfockb}
\end{equation}
Using definitions of the one-electron and one-hole projector,  
\eqs{eq:densmat_elecd} and \reff{eq:densmat_hole} respectively,
the usual expression of the first-order linear-response of the density 
matrix is recovered.\cite{diercksen_jcp_1966,moccia_tca_1967,dodds_molphys_1977-1,
dodds_molphys_1977-2,wolinski_molphys_1980,wolinski_jacs_1990} 
In operator form, it gives
\begin{align}
\hat{D}^{(1)} &= \sum^{}_{i\in\trm{occ}}\sum^{}_{j\in\trm{virt}}
	\frac{\langle\psi_i|\hat{H}^{(1)}|\bar{\psi}_j\rangle}{\Em_i - \Eb_j}\nonumber
|\psi_i\rangle\langle\bar{\psi}_j| \\
&+  \trm{conjugate transpose}
	  \label{eq:sosfockb}
\end{align}
Assuming that the sets of  non-perturbed eigenvectors, 
$\{\psi_i\}_{i=1}^{\Ne}$
and $\{\bar{\psi}_j\}_{j=1}^{\Nb}$, 
are properly orthonormalized, then, the fact that
$\trace{|\psi_i\rangle\langle\bar{\psi}_j|}=\langle\bar{\psi}_j|\psi_i\rangle=0\;\forall (i,j)$, 
guarantee the $N$-representability conditions of \eqs{eq:idemexp0} and \reff{eq:idemexpa}

Derivation for the second- and third-order linear-responses are given in the Appendix \ref{app:sos-dmpt}.
From here, we shall briefly review the generalized working equations needed for solving
the density matrix response at any order $k>1$. The off-diagonal contributions are given by
\begin{eqnarray}
	\Dmovk{k} &=& \sum^{}_{i\in\trm{occ}}\sum^{}_{j\in\trm{virt}}
	\frac{\Fmovijk{k} - \sum_{l=1}^{k-1}[\Fmk{l},\Dmk{k-l}]_{ov,ij}}{\Em_i - \Eb_j} \label{dbov1_dcpscf}\\
	&=& \sum_{l=1}^{k}\sum^{}_{i\in\trm{occ}}\sum^{}_{j\in\trm{virt}}
	\frac{[\Dmk{k-l},\Fmk{l}]_{ov,ij}}{\Em_i - \Eb_j}
	\label{dbov2_dcpscf}
\end{eqnarray}
In operator form it gives:
\begin{equation}
	\hat{D}_{ov}^{(k)} = \sum^{k}_{l=1}\sum_{i\in\trm{occ}}\sum_{j\in\trm{virt}}\frac{\langle\psi_{i} | 
	    [\hat{D}^{(k-l)},\hat{H}^{(l)}]|\bar{\psi}_j\rangle}{\epsilon_{i}-\bar{\epsilon}_{j}} |\psi_i \rangle\langle \bar{\psi}_{j} |
\label{dbov3_dcpscf}	
\end{equation}
The diagonal terms are given by
\begin{eqnarray}
	\Dmook{k}  =  - \Dm \left( \sum\limits_{l=1}^{k-1} \Dmk{l}\Dmk{k-l} \right) \Dm 
	\label{dboo_dcpscf}\\
	\Dmvvk{k}  =  + \Db \left( \sum\limits_{l=1}^{k-1} \Dmk{l}\Dmk{k-l} \right) \Db
	\label{dbvv_dcpscf} 
\end{eqnarray}    
Again here, given a set of orthonormalized non-perturbed eigenvectors,
it is easy to show that, $\trace{\Dmovk{k}}=0=\trace{\Dmook{k}+\Dmvvk{k}}$,
ensuring the respect of the generalized perturbed $N$-representability conditions 
of \eq{eq:idemexpg}. Note that for a Hamiltonian perturbation expansion up to
the 1st order in \eq{eq:Fmpert} only the commutator $[\Dmk{k-1},\Fmk{1}]$
survives in the $\sum_l$ of \eq{dbov2_dcpscf}.

Although, McWeeny's formulation of DMPT is based on the density-matrix
it still requires the knowledge of the unperturbed eigenstates which means that
at least one Hamiltonian diagonalization must be performed prior entering the 
DMPT resolution. As it shall be shown below, there exist alternative solutions which
allow us to circumvent the expensive diagonalization step.

\subsection{Sylvester-DMPT}
\label{sec:cg-dcpscf}

As in the unperturbed case, to completely bypass calculation of the eigenstates 
when solving the DMPT equations, an objective functional with perturbed density 
matrices as degree of freedom has to be defined, and minimized without the support 
of the spectral decomposition \reff{eq:mcweenyfocka}. In this respect, Ochsenfeld and 
Head-Gordon have proposed to extend the of LNV functional\cite{LNV_PRB_1993} minimization 
principle to DMPT.\cite{ochsenfeld_cpl_1997} 
Later, Kussmann and Ochsenfeld reformulated the working equations to cure for 
numerical instabilities.\cite{kussmann_thesis_2006,kussmann_jcp_2007-1} 
The approach relies on solving \eq{eq:scfexpg} subject to 
commuting with the unperturbed density matrix. 
For instance, at the first-order of perturbation, on multiplying \eqs{eq:scfexpa} 
from the left and from the right  by $\Dm$ separately, and substracting, 
we obtain\footnote{The passing from $[\Dm,[\Fm,\Dmk{1}]]+[\Dm,[\Fmk{1},D]=0$
to \eq{eq:dcpscf0} required the use of the identity: $[A,[B,C]] = [B,[A,C]] + [C,[B,A]]$.}
\begin{equation}
	[ \Fm, [ \Dm, \Dmk{1} ] ] + [ \Dmk{1}, [ \Dm, \Fm ] ] +  [ \Dm, [\Fmk{1}, \Dm ] ] = 0
\label{eq:dcpscf0}
\end{equation}
Since we assume that the exact zero-order density matrix is known, the second term in \eq{eq:dcpscf0}
vanishes. By noting that:
\begin{equation}
	[ \Dm, [\Fmk{1}, \Dm ] ] = 2\Dm\Fmk{1}\Dm - \lbrace \Dm^2, \Fmk{1} \rbrace 
\label{eq:dcpscf_}
\end{equation}
with $\{\cdot,\cdot\}$ the symbol for the anticommutator of two operators,
then, \eq{eq:dcpscf0} simplifies to:
\begin{equation}
	[ \Fm, [ \Dm, \Dmk{1} ] ] =  
	\lbrace \Dm, \Fmk{1} \rbrace -  2 \Dm\Fmk{1}\Dm 
\label{eq:dcpscf1}
\end{equation}
A practical form for solving \eq{eq:dcpscf1} is obtained by expanding the commutators 
of the lhs. Using the identity of \eq{eq:idemexpa}, this yields to
\begin{equation}
(2\Fm\Dm - \Fm)\Dmk{1} + \Dmk{1}(2\Dm\Fm - \Fm) 
= \lbrace \Dm, \Fmk{1} \rbrace -  2 \Dm\Fmk{1}\Dm 
\label{eq:dcpscf1_1}
\end{equation}
which can be identified as being a Sylvester-like equation of kind $AX+XB=C$,\cite{brandts_2001,simoncini_computational_2016} 
where $B:=A^t$, and $X:=\Dmk{1}$ is the unknown to solve for.\footnote{Note 
that the special case where $B=A^t$ in the Sylvester equation $AX+XB=C$ is also referred to  
as the Lyapunov equation in litterature\cite{simoncini_computational_2016}}
It is worthwhile to note that on multiplying \eq{eq:dcpscf1} from the left by $\Dm$, 
and from the right by $\Db$, and conversely, by $\Db$ on the left and $\Dm$ on the right, 
\eqs{eq:mcwfstorder} are recovered. By induction, the DMPT equation \reff{eq:dcpscf1_1}
can be generalized to any order $k>1$, with the \textit{k}th-order transition 
matrix $\Dmk{k}$ solution of
\begin{align}
	\left[ \Fm,[ \Dm, \Dmk{k} ] \right] 
	&= \lbrace \Dm, \Fmk{k} \rbrace - 2 \Dm\Fmk{k}\Dm \nonumber\\
	&+ \sum\limits_{l=1}^{k-1}\left[ \Dm, [ \Dmk{k-l}, \Fmk{l} ] \right] \label{eq:dcpscf_k}
\end{align} 
As for the first order, by expanding the commutators and using identity of \eq{eq:idemexpg} we found
\begin{widetext}
\begin{equation}
(2\Fm\Dm - \Fm)\Dmk{k} + \Dmk{k}(2\Dm \Fm- \Fm)
= \lbrace \Dm, \Fmk{k} \rbrace -  2 \Dm\Fmk{k}\Dm
+ \sum\limits_{l=1}^{k-1} \left[ \Dm, [ \Dmk{k-l}, \Fmk{l} ] \right] - \left\{\Dmk{l}\Dmk{k-l},\Fm\right\}
\label{eq:dcpscf_k_1}
\end{equation} 
\end{widetext}
Assuming that all the lower order (up to $k-1$) perturbed density matrices are known
in the rhs of the equation, then $\Dmk{k}$ in the lhs can be found by solving 
the Sylvester equation
\begin{equation}
	AX + XA^t = C 
\label{linear_dcpscf} 
\end{equation}    
where $(A,X,C)\in\real^{M\times M}$ are square matrices. The fixed matrix $A$ 
corresponds to $(2\Fm\Dm-\Fm)$ and the fixed matrix $C$ is given by the rhs of \eq{eq:dcpscf_k_1}.
In this work, the algorithm of Bartels and Stewart\cite{bartels_solution_1972} (BS)
shall be used to solve \eq{linear_dcpscf}.
Alternatives to the BS algorithm are envisageable by vectorizing \eq{linear_dcpscf} which 
transforms the Sylvester equation to a standard linear system of equations.
Among the numerous iterative methods developed for solving linear system of equations,\cite{saad_iterative_2003}
the conjugate-gradient (CG) minimization is one the most efficient,\cite{nocedal_numerical_2006} 
especially for large scale problems presenting a sparsity pattern. After a set of trials on model
systems (cf. Sec.~\ref{sec:results}), we found that the BS algorithm was more efficient than
CG minimization.

Note that the CG method is also a popular alternative to iterative diagonalizations, \eg\ the Davidson method,\cite{davidson_iterative_1975}
when it is sufficient to access to the partial set of the $\tilde{M}$ lowest energy eigenstates, with $N\leq\tilde{M}<M$,
as for instance for non-vanishing band gap system, where only the subset of occupied states are required 
to build $\Dm$ in \eq{eq:densop_idemp}. This fact is exploited by iterative diagonalizations when the number of basis 
functions exceeds by far $N$, the typical case being the planewave (PW) basis set. For insulators,
the approximate first $N$ lowest eigenstates in some Krylov subspace of $\Fm$ suffices to achieve 
the desired accuracy.\cite{wood_new_1985,kresse_efficient_1996} 
In this context, band-by-band (also called state-by-state) CG algorithms (BB-CG),\cite{teter_solution_1989,
payne_iterative_1992,kresse_efficient_1996,giannozzi_quantum_2009} which basically 
perform sequential CG minimizations under some orthormalization constraints, 
have also proved to be a valuable alternative.\cite{kresse_efficiency_1996} 

The BB-CG method is one of the ingredient of the PW-based (density functional) perturbation 
theory.\cite{gonze_adiabatic_1995,gonze_first-principles_1997,baroni_greens-function_1987,baroni_phonons_2001} 
In constrast to the McWeeny-DMPT, which requires the knowledge of the full eigenspectrum of $\Fm$,
it is possible to compute any of the $k$th-order density matrix, using the sole information available 
from the occupied eigenstates. This constitutes the framework of the high-order DMPT
introduced by Lazzeri and Mauri.\cite{lazzeri_first-principles_2003,lazzeri_high-order_2003} 
The strong overlap existing between this approach, the McWeeny- and the Sylvester-DMPT 
is discussed in Appendix \ref{app:lm-dmpt}. 

\subsection{Purification-DMPT}
\label{sec:rp-dcpscf}

A powerful alternative to the McWeeny- and Sylvester-DMPT resides in the
density matrix purification method,\cite{niklasson_density_2011} which, at the zero order, 
given the unperturbed Hamiltonian matrix, consists of finding the corresponding ground state density matrix
by approaching the Heaviside step function of \eq{eq:fermidirac_idemp} using a polynomial recursion.  
Within the canonical ensemble, it can be formally expressed as:
\begin{subequations}
\label{eq:pr_algo}
\begin{align}	
	\Dm_0        &= \Ipoly{P}(\Fm;\{\Ne,\ldots\}) \label{eq:pr_algo_a}\\
	\Dm_{n+1} &= \Kpoly{P}(\Dm_n;\{\Ne,\ldots\})\label{eq:pr_algo_b}\\
	\trm{such that}&\quad\Dm_{\infty} = \lim_{n\rightarrow\infty} \Dm_n\label{eq:pr_algo_c}
\end{align}	
\end{subequations}
where $\Dm_{0}$, $\Dm_{n+1}$ and $\Dm_{\infty}$ designate the initial, 
the $(n+1)$-iterate, and the converged density matrix, respectively.
The polynomial recursive sequence is initiated by a linear mapping 
\reff{eq:pr_algo_a}, where the function $\Ipoly{P}$ rescales, shifts 
and reverses the eigenspectrum of $\Fm$ into the proper interval 
for occupation numbers, \ie\ $\forall\:i:\rho_i\in[0,1]$. In that case, 
the initial guess, $\Dm_0$, represents some ground state in the 
sense of \eq{eq:densop}.
Then, by applying recursively the polynomial function, $\Kpoly{P}$,
the degenerate sets of $\Ne$ and $\Nb$ eigenvalues of $\Dm_n$  associated 
with the occupied and virtual subspaces are progressively brought towards 1 and 0, respectively.
These subsets will be symbolized by: $\{o\}\equiv\{\nocc_i| \lim_{n\rightarrow\infty}\nocc_i=1\}_{i=1}^{\Ne}$ 
and $\{v\}\equiv\{\nocc_i | \lim_{n\rightarrow\infty} \nocc_i=0\}_{i=\Ne+1}^M$.
At convergence, $\Dm=\Dm_{\infty}$ such that $\Dm$ fulfils the 
$N$-representability conditions \reff{eq:nrep_idemp} and the ground-state 
occupation \reff{eq:fermidirac_idemp} at $T=0$, without prior knowledge of the chemical 
potential. 

In this work, we have considered two different purification schemes, the second-order 
trace-correcting (TC2) purification\cite{Niklasson_TCP_2002} ---later rebaptised the 
second-order spectral projection\cite{Cawkwell_JCTC_2012}--- and the trace-conserving 
hole-particle canonical purification\cite{truflandier_jcp_2016} (HPCP). 
The original TC2 recursive polynomial\cite{Niklasson_TCP_2002,niklasson_jcp_2005} 
is given by
\begin{subequations}
\label{eq:tc2_rec}
\begin{align}	
	\Kpoly{\trm{TC2}}(\Dm_n;\{\Ne\}) &= \Dm_n 
	+ 2\left(\Theta(\Delta N_n)-\frac{1}{2}\right)\Dm_n\bar{\Dm}_n\label{eq:tc2_reca} \\
	\trm{with}\quad\Delta N_n &=\Ne-\trace{\Dm_n}\nonumber\\
	\quad\trm{such that}\quad&\lim_{n\rightarrow\infty}\Delta N_n=0\label{eq:tc2_recb}\\
	\trm{and}\quad\Theta(x)    &= 
	\left\{ 
		\begin{array}{cc}
		0   & \trm{if}\quad x\leq 0\\ 
		1   & \trm{if}\quad x   >  0\\
		\end{array}
	\right. \label{eq:tc2_recc}
\end{align}	
\end{subequations} 
along with the following initialization mapping,
\begin{equation}
	\Ipoly{\trm{TC2}} (\Fm) = \frac{\etmax\Id - \Fm}{\etmax - \etmin}\label{eq:tc2_init}\\
\end{equation}
In the above equation, $(\etmin, \etmax)$ are approximated values of the lower 
and upper bounds of the Hamiltonian matrix eigenspectrum $(\emin, \emax)$. They can 
be easily estimated, \ie\ without the support of iterative diagonalization, from the 
Ger{\v s}gorin's disc theorem\cite{Gerschgorin_disc_1931} with the following 
convenient properties: $\etmin<\emin$ and $\etmax>\emax$. It should be emphasized
that the initial guess generated from \eq{eq:tc2_init} is not $N$-representable 
since $\trace{D_0}$ is not constrained to be equal to $\Ne$ ---indeed it must not be.
As evidenced by \eq{eq:tc2_recc}, the TC2 purification is to be regarded as a discontinuous self-mapping 
of $\Dm_n$, where the sign\footnote{Even if obvious, we recall that: $\sign{(x)}=2\Theta(x)-1$. 
The TC2 formulation of \eq{eq:tc2_reca} using $\Theta(x)$ instead of the $\sign(x)$ 
is only useful when compared with the HPCP polynomial \reff{eq:hpcp_rec}.} of the deviation 
of $\trace{D_n}$ with respect to $\Ne$ dictates the polynomial to apply at iteration $n+1$.
If at the current state $\trace{D_n}<N$, all the occupation numbers, but those already trapped 
at the fixed points 0 and 1, will be moved (at different rates ; the closer to 1/2 the faster) 
towards the turning point $x_p=1$. Conversely, for $\trace{D_n}>N$, the occupation
numbers are moved towards $x_{\bar{p}}=0$.
In principle, the recursive sequence should be terminated when $|\Delta N_n|$ or/and 
another relevant convergence criteria is/are below some threshold value,\cite{niklasson_density_2011} 
\eg\ $||\Dm_n\Db_n||$. As a matter of fact, the Heaviside 
singularity ---especially for vanishing-gap systems--- may pose some difficulties in achieving 
proper convergence, since in the definition \reff{eq:tc2_recc} there is no fixed midpoint 
for $\Theta(x=0)$. It is likely the cause of numerical unstabilities when approaching convergence.
In this respect, to cure possible issues of the TC2 purification, 
several refinements of \eq{eq:tc2_recb} ---and the expression of the stopping criteria--- 
have been proposed.\cite{niklasson_density_2011,Cawkwell_JCTC_2012,Cawkwell_JCTC_2014} 
These refinements, which are more substantial when sparse linear algebra is 
applied,\cite{weber_jcp_2005,niklasson_jcp_2005,niklasson_jcp_2007} 
can lead to a significant increase of the algorithmic complexity.\cite{rubensson_siam_2012,
rubensson_jcp-3_2008}

Instead of applying a global upward or downward shift to all the non-converged $\{\nocc_i\}$,
one can seek to simultaneously operate on $\{o\}$ and $\{v\}$ during the
polynomial recursion. The easiest way is to increase the polynomial degree from 2 to 3, 
in order to introduce an inflexion point, $\xflex\in]0,1[$, separating the two subsets.
For instance, assuming a symmetric model, when half of the available states are occupied,
\ie\ the filling factor $\theta=N/M$ equals $1/2$, and the chemical potential is at the 
midpoint of $[\emin,\emax]$, $\xflex=1/2$ is the optimal position.
In this case, all the $\nocc_i\in\{o\}$ verifying $\nocc_i>\xflex$ are push towards 1, 
whereas at the same time, all the $\nocc_i\in\{v\}$ verifying $\nocc_i<\xflex$ are push 
towards 0. By recognizing the constant $1/2$ in \eq{eq:tc2_reca} as $\xflex$, 
and on substituting $\Theta$ by $\Dm$ in the same equation, it is intuitively 
easy enough to see that the resulting polynomial of degree 3 fulfils the requirements 
stated above. When compared to the TC2 polynomial, both, $x_p$ and $x_{\bar{p}}$,
are now stable fixed points, and more importantly the iterative mapping is continuous.
Actually, it does correspond to the well-known McWeeny recursive sequence,\cite{McWeeny_density_1956a}
\begin{align}
	\Kpoly{\trm{McW}}(\Dm_n) &= 
	\Dm_n + 2\left(\Dm_n-\frac{1}{2}\Id\right)\Dm_n\bar{\Dm}_n\nonumber\\
	&=3\Dm_{n}^2-2\Dm_{n}^3 \label{eq:mcw_rec}
\end{align} 
Unfortunatly, in the general case, unconstrained application of \eq{eq:mcw_rec} 
is likely to deliver $\Dm_{\infty}$ verifying \eq{eq:idemp} but with 
$\trace{\Dm_{\infty}}\neq\Ne$, especially when $\theta$ is far from $1/2$. 
Given any $(\Fm,\Ne,\N)$, if we search for generalizations of the McWeeny purification, 
two difficulties must be addressed: (i) how to dynamically adapt $\xflex$ to 
non-symmetric $\{\nocc_i\}$ distributions while maintaining the two stable (un)fixed 
points (nearby) at 0 and 1 ; (ii) how to ensure that the converged density 
matrix is $\Ne$-representable? Although they are not all mandatory,\cite{} 
we can consider 3 constraints under which the problem shall be solved: (a) 
no \textit{a priori} information on the structure of $\Fm$ eigenspectrum, 
nor on some of the interior eigenvalues are required, (b) the highest polynomial 
degree is 3, and (c) the recursive mapping must remain continuous. A first solution 
was brought by Palser and Manolopoulos (PM) in their $\textit{NVT}$ version of 
the McWeeny purification.\cite{PM_PRB_1998} 
Recently, by solving a constrained optimization problem dealing with the 
idempotency error minimization of $\Dm$,\cite{truflandier_jcp_2016} 
we found that the PM polynomial could be significantly simplified and 
accelerated through the hole-particle duality.\cite{Mazziotti_PRE_2003}
The resulting hole-particle canonical purification (HPCP) polynomial\cite{truflandier_jcp_2016} 
is given by:
\begin{subequations}
\label{eq:hpcp_rec}
\begin{align}
	\Kpoly{\trm{HPCP}}(\Dm_n)  &= \Dm_{n} + 2 \left( \Dm_n - c_n\Id\right)\Dm_{n}\Db_{n}\label{eq:hpcp_rec_a}\\
	\trm{with}\quad c_n &= \frac{\trace{\Dm^2_n\Db_n}}{\trace{\Dm_n\Db_n}}\\
	\quad\trm{such that}\quad&\lim_{n\rightarrow\infty}c_n=\frac{1}{2}\label{eq:hpcp_rec_b}
\end{align}	
\end{subequations} 
with the linear mapping function,
\begin{subequations}
\label{eq:hpcp_init}
\begin{align}
	   \Ipoly{\trm{HPCP}} (\Fm;\{N,M,\alpha\}) &= \left(\alpha\beta_{\trm{min}} + (1-\alpha)\beta_{\trm{max}}\right)\nonumber\\
	   &\times\left(\tilde{\mu}\Im-\Fm\right)+\theta\Im \\
	   \trm{with}\quad
	   \tilde{\mu} & =\frac{\trace{\Fm}}{\N}\\
	   \beta_{\trm{min}} = \min&\left( \frac{\theta }{\etmax-\tilde{\mu}},	
	   \frac{1-\theta}{\tilde{\mu}-\etmin}\right)\\
	   \beta_{\trm{max}}= \max&\left( \frac{\theta }{\etmax-\tilde{\mu}},	
	   \frac{1-\theta}{\tilde{\mu}-\etmin}\right)
\end{align}	
\end{subequations}
where $\alpha$ is a mixing parameter which can be optimized with
respect to $\theta$.\cite{truflandier_jcp_2016} For the numerical 
experiments presented in \sec{sec:results}, $\alpha$ shall be fixed to $1/2$.
As in the original McWeeny purification, the HPCP polynomial presents
an unstable fixed point, $c_n$, which in the present case, modulates 
the position of $\xflex$ and enforces the correction term in the rhs 
of \eq{eq:hpcp_rec} to be traceless. Provided a $N$-representable 
$\Dm_0$ using \eqs{eq:hpcp_init}, the HPCP  is capable of converging
self-consistently to the ground-state density matrix while verifying the 
$N$-representability conditions throughout the purification process.
Note that, when approaching convergence, it can be shown\cite{PM_PRB_1998} 
that $c_n\rightarrow 1/2$. Nevertheless, in this regime, the numerator 
and denominator of \eq{eq:hpcp_rec_b} tends to zero. As a result, to 
avoid numerical unstabilities related to floating-point round-off, it is safer 
to fix $c_n$ to $1/2$ at the very late stage of the purification.

It must be stressed that the TC2 and HPCP methods are very distinct
in their \textit{NVT} minimization principle. For the HPCP, $\Ne$ is kept 
fixed while $T$ is implicitly minimized ---to zero for non-vanishing 
band-gap system---, whereas for the TC2 purification, $T$ is implicitly fixed 
to zero, and $\Ne$ is perturbatively optimized around the target value, 
resulting in very different convergence profiles regarding for instance
monotonicity and variational properties (\textit{vide infra}).%

It is rather remarkable that the perturbed density matrices involved
in the power series of $\Dmp$ can also be determined by a straight 
application of the polynomial recursive sequence \reff{eq:pr_algo} 
with $\Fm_{\lambda}$ as input.\cite{niklasson_prl_2004}
The perturbed analogue of \eq{eq:pr_algo} formally writes:
\begin{subequations}
\label{eq:pr_algo_pert}
\begin{align}	
&\trm{initialisation:}\: \left( 
\begin{array}{c}
\Dmk{0}_{0}\\
\Dmk{1}_{0}\\
\vdots  \\ 
\Dmk{k}_{0}  
\end{array}
\right) 
=
\left( 
\begin{array}{c}
\Ipolypertk{0}{P}(\Fmk{0})\\
\Ipolypertk{1}{P}(\Fmk{1})\\
\vdots \\ 
\Ipolypertk{k}{P}(\Fmk{k})
\end{array}%
\right)\label{eq:pr_algo_pert_a} \\
&\trm{recursion:}\: \left( 
\begin{array}{c}
\Dmk{0}_{n+1}\\
\Dmk{1}_{n+1}\\
\vdots  \\ 
\Dmk{k}_{n+1}
\end{array}
\right) 
=
\left( 
\begin{array}{c}
\Kpolypertk{0}{P}(\Dmk{0}_n)\\
\Kpolypertk{1}{P}(\Dmk{0}_n,\Dmk{1}_n)\\
\vdots \\ 
\Kpolypertk{k}{P}(\Dmk{0}_n,\Dmk{1}_n,\ldots\Dmk{k}_n)
\end{array}%
\right)\label{eq:pr_algo_pert_b}\\
&\trm{such that:}\:\nonumber\\
&\Dm_{\infty},\Dmk{1}_{\infty},\ldots,\Dmk{k}_{\infty}:= 
\lim_{n\rightarrow\infty}\left(\Dm_{n},\Dmk{1}_{n},\ldots,\Dmk{k}_{n}\right) \label{eq:pr_algo_pert_c}
\end{align}	
\end{subequations}
where $\Ipolypertk{0}{P}$ and $\Kpolypertk{0}{P}$ correspond to the unperturbed
linear mapping and recursive polynomial as introduced in Eqs.~\reff{eq:pr_algo_a} and \reff{eq:pr_algo_b}, respectively.
Eqs~\reff{eq:pr_algo_pert} outline that, provided some set of 0-to-$k$ order Hamiltonian matrices
to initialise the 0-to-$k$ order density matrices, repeated application of 
the $k$-perturbed recursive sequence deliver better estimates of the $(k+1)$ input 
quantities by propagating the perturbed purification. 
Note that, evaluation of the higher order perturbed term does
not involve prior exact knowledge of the lower orders ---not even the exact unperturbed density matrix---, 
instead all the orders are resolved on-the-fly.
Compared to the standard approaches, \eg\ the McWeeny- and Sylvester-DMPT
which are based on solving order-by-order $(k+1)$-sets of linear equations
with $\Dmk{k}$ as unknown, the purification-DMPT performs the resolution of one 
unique non-linear equation with $(k+1)$ unknowns.
In order to avoid heavy notations, the $(k+1)$-dimension of the inputs and outputs
in Eqs.~\reff{eq:pr_algo_pert} will be simplified by retaining only the $k$-order term in the function
argument and value, respectively.

By substituting \eq{eq:Fmpert} into \eq{eq:tc2_init}, and \eq{eq:Dmpert} 
into \reff{eq:tc2_reca}, equating the perturbation orders, the  
$k$-perturbed component of the TC2 recursive sequence\cite{weber_prl_2004} 
writes
\begin{subequations}
\label{eq:tc2_dmpt_poly}
\begin{align}
	\Ipolypertb{\trm{TC2}}(\Fmk{k})               &=  
	-\frac{\Fmk{k}}{\etmax - \etmin} \label{eq:tc2_dmpt_rec_a} \\	
	\Kpolypert{\trm{TC2}}(\Dmk{k}_n;\{\Ne\}) &= \nonumber\\
	\Dmk{k}_n &+ 2\left(\Theta(\Delta N_n)-\frac{1}{2}\right)\sum^{k}_{l=0}\Dmk{l}_n\Dbk{k-l}_n \label{eq:tc2_dmpt_rec_b}
\end{align}	 
\end{subequations} 
By referring to \eqs{eq:idemexp}, it is easily seen that the sum over the perturbed 
hole-particle density matrix product appearing in the rhs of \eq{eq:tc2_dmpt_rec_b} corresponds to 
the error in the idempotency (noted $\Delta\Dmk{k}$ below). On remembering that $\Dbk{0}=\Id - \Dmk{0}$ 
and $\Dbk{k>0}=-\Dmk{k>0}$, such that, at iterate $n$,
\begin{subequations}
\label{eq:error_idemp}
\begin{align}
	\Delta\Dmk{0}_n &= \Dm_n - \Dm_n^2\label{eq:error_idempa}\\
	\Delta\Dmk{1}_n &= \Dmk{1}_n - 
	\left(\Dm_{n}\Dmk{1}_n +  \Dmk{1}_{n}\Dm_{n}\right)\label{eq:error_idempb}\\
	\Delta\Dmk{2}_n &= \Dmk{2}_n - 
	\left(\Dm_{n}\Dmk{2}_n + (\Dmk{1}_{n})^2 + \Dmk{2}_{n}\Dm_{n}\right)\label{eq:error_idempc}\\
	\vdots \nonumber\\
	\Delta\Dmk{k}_n &= \Dmk{k}_n - \sum^{k}_{l=0} \Dmk{l}_{n}\Dmk{k-l}_{n}\label{eq:error_idempd}	
\end{align}	 
\end{subequations} 
we obtain the more compact TC2 perturbed recursion formula
\begin{equation}
\label{eq:tc2_dmpt_rec}
	\Dmk{k}_{n+1} \overset{\mathrm{TC2}}{=} \Dmk{k}_n + 2\left(\Theta(\Delta N_n)-\frac{1}{2}\right)\Delta\Dmk{k}_n
\end{equation}
By proceeding the same way with the hole-particle canonical purification 
initialization and recursive polynomial ---\eqs{eq:hpcp_init} and \reff{eq:hpcp_rec} respectively---, 
the $k$-perturbed component of the HPCP recursive sequence writes
\begin{subequations}
\label{eq:hpcp_dmpt_poly}
\begin{align}
	\Ipolypertb{\trm{HPCP}}(\Fmk{k};\{\Ne\}) &=  
	-\left(\alpha\beta_{\trm{min}} + (1-\alpha)\beta_{\trm{max}}\right)\Fmk{k} \label{eq:hpcp_dmpt_poly_a} \\	
	\Kpolypert{\trm{HPCP}}(\Dmk{k}_n) &= \Dmk{k}_n \nonumber\\
	+ 2\sum^{k}_{l=0}&\left(\sum^{l}_{m=0}\Dmk{m}_n\Dmk{l-m}_n - c_n\Dmk{l}_n\right)\Dbk{k-l}_n
	\label{eq:hpcp_dmpt_poly_b}
\end{align}	 
\end{subequations} 
Using definitions \reff{eq:error_idemp}, after some rearrangement, we find
\begin{equation}
\label{eq:hpcp_dmpt_rec}
	\Dmk{k}_{n+1} \overset{\mathrm{HPCP}}{=} \Dmk{k}_n 
	+ 2\left(\Dm_n - c_n\Id\right)\Delta\Dmk{k}_n
    + 2\sum^{k-1}_{l=0}\Dmk{k-l}_n\Delta\Dmk{l}_n
\end{equation}
It should be stressed that since the unstable fixed-point in \eq{eq:hpcp_dmpt_rec} depends on
the unperturbed density matrix, resolving $\Dmk{k}$ order-by-order is not possible with the current formalism. 
A full decoupling of the perturbed purification would require to establish and solve a constrained 
optimization problem as in \citen{truflandier_jcp_2016}, with the $k$-order perturbed idempotency 
relations \reff{eq:idemexp} as the main ingredient for the quadratic functional form to minimize. 
This possibility shall be addressed in a further study.

Comparing \eq{eq:hpcp_dmpt_rec} to (\ref{eq:tc2_dmpt_rec}), 
it is clear that the perturbed HPCP approach involves additional correction
terms in the polynomial expansion, increasing its computational
complexity with respect to the perturbed TC2. Looking at the 
most computationally demanding task,\footnote{Maybe
it is worth to recall that for square matrices of size $M\times M$,
the number of floating-point operations (FLOPs) for dense matrix-matrix 
multiplication scales as $O(M^3)$, compare to $O(M^2)$ for all 
the other matrix operations involved in those perturbed recursions, 
\eg\ scalar multiplication, addition and transposition.}
\ie\ the number of matrix multiplications (MMs), which in 
both cases increases linearly with the perturbation order, 
it is found that the perturbed HPCP requires nearly 3 times 
more MMs than the perturbed TC2.\footnote{For $k$ odd, 
exactly $(k+1)/2$ MMs must be performed for the TC2
\vs\ $(3k+3)/2$ for the HPCP. For $k$ even, it is found
$(k+2)/2$ \vs\ $(3k+4)/2$ MMs, respectively.} 
This is balanced out by the rate of convergence, 
which, as in the unperturbed case, is linear for the TC2, but quadratic 
for the HPCP. 

\section{Examples and performances}
\label{sec:results}

To illustrate the perturbation-based DMPT, we have considered two
examples taken from the $\pi$-bonding perturbation in H{\"u}ckel theory.\cite{orbitals_1985}
Our aim here is to test the purification approaches, both in terms of stability
and convergence.
The first one consists of decomposing the H{\"u}ckel matrix of the benzene molecule into (\textit{i}) a 
non-perturbed Hamiltonian containing the matrix elements of the butadiene and ethylene 
subunits, and (\textit{ii}) a first-order perturbed Hamiltonian associated with their coupling. 
The decomposition is explicitly given below:
\begin{align}	
H_{\lambda}(\trm{benzene}) = \Fmk{0} &+ \lambda\Fmk{1} =   \nonumber\\
\left( 
\begin{array}{cccccc}
\alpha  & \beta  &      0   & 0        & 0         & 0   \\
\beta   & \alpha & \beta  &  0       & 0         & 0   \\
0         & \beta  & \alpha & \beta  & 0         & 0   \\
0         & 0        & \beta  & \alpha & 0         & 0    \\
0         & 0        &    0     & 0        & \alpha  & \beta \\
0         & 0        &    0     & 0        & \beta   & \alpha \\
\end{array}
\right) 
&+
\lambda
\left( 
\begin{array}{cccccc}
0                       & 0                      & 0                    & 0                      & 0                        & \beta\\
0                       & 0                      & 0                    & 0                      & 0                        & 0                      \\
0                       & 0                      & 0                    & 0                      & 0                        & 0                      \\
0                       & 0                      & 0                    & 0                      & \beta                  & 0                       \\
0                       & 0                      & 0                    & \beta                 & 0                       & 0                       \\
\beta                 & 0                      & 0                    & 0                       & 0                       & 0                       \\
\end{array}
\right)  \nonumber
\end{align}
where $\alpha$ and $\beta$ are the usual symbols for the carbon Coulomb and resonance integrals, respectively.
We note that in this example the perturbation matrix is purely non-diagonal. For the second example, we have chosen
the H{\"u}ckel matrix of the pyridine molecule for which one of the $sp^2$-C in benzene is substituted by a 
$sp^2$-N atom. Considering the benzene H{\"u}ckel matrix as the zero-order Hamiltonian, the perturbation 
matrix contains the variation of the Coulomb ($\Delta\alpha$) and resonance ($\Delta\beta$) integrals related to 
the nitrogen/carbon substitution. The corresponding total Hamiltonian writes:
\begin{align}	
H_{\lambda}(\trm{pyridine}) = \Fmk{0} &+ \lambda\Fmk{1} =  \nonumber\\
\left( 
\begin{array}{cccccc}
\alpha  & \beta   &     0    & 0        & 0         & \beta\\
\beta   & \alpha  & \beta  &  0       & 0         & 0     \\
0         & \beta   & \alpha  & \beta & 0         & 0     \\
0         & 0         & \beta   & \alpha & \beta  & 0      \\
0         & 0         &    0      & \beta  & \alpha  & \beta \\
\beta   & 0        &    0      & 0         & \beta   & \alpha \\
\end{array}
\right) 
&+
\lambda\left( 
\begin{array}{cccccc}
\Delta\alpha   & \Delta\beta       & 0                    & 0                      & 0                       & \Delta\beta\\
\Delta\beta    & 0                      & 0                    & 0                      & 0                       & 0                      \\
0                   & 0                      & 0                    & 0                      & 0                       & 0                      \\
0                   & 0                      & 0                    & 0                      & 0                       & 0                       \\
0                   & 0                      & 0                    &0                       & 0                       & 0                       \\
\Delta\beta    & 0                      & 0                    & 0                      & 0                       & 0                       \\
\end{array}
\right) \nonumber
\end{align}
with, for nitrogen, the following parameters: $\Delta\alpha=\beta/2$ and $\Delta\beta=-0.2\beta$. 
For numerical experiments we set $\alpha$ and $\beta$ to -11.400 and -2.568 eV, respectively.
Stopping criteria for purification was based on the norm of unperturbed density matrix iterates 
such that the purification process was stopped for $||D_{n+1}-D_{n}||_{\mathcal{F}}<10^{-12}$.
Further tests were performed on the Frobenius norms and traces of all the perturbed 
density matrices to ensure full convergence at all the orders. 

\begin{widetext}
\begin{figure}[h!]
\centering\includegraphics[width=0.95\textwidth]{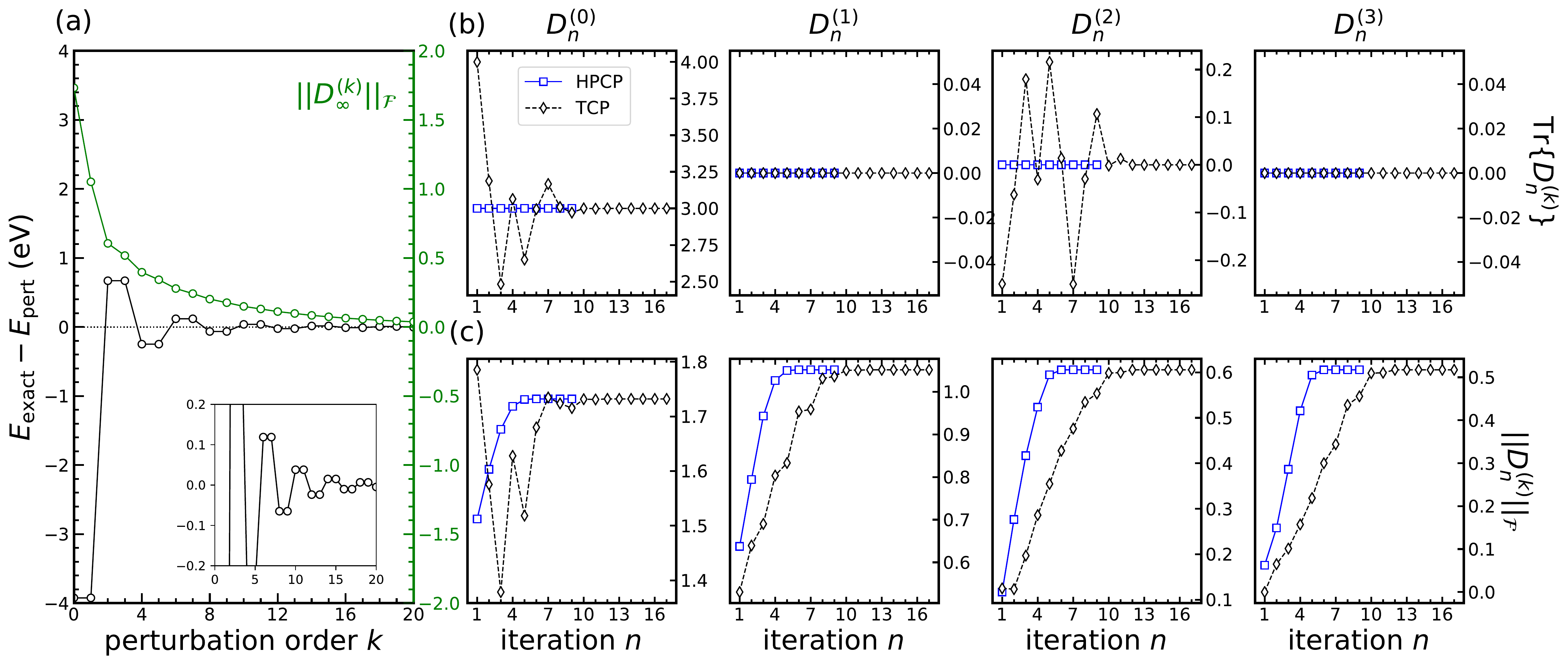} 
\caption{Perturbation series for $\pi$-bonding in benzene using perturbed H{\"u}ckel matrices (see text for details). 
(a) Convergence of the perturbed energy and Frobenius norm of the corresponding perturbed density 
matrices using purification-DMPT. The convergence of the unperturbed and firsts 3 perturbed 
density matrices during the purification cycles are presented in terms of their traces (b), and their 
norms (c), for both the TC2 and HPCP polynomials.}
\label{fig:benzene}
\end{figure}
\end{widetext}
\begin{widetext}
\begin{figure}[h!]
\centering\includegraphics[width=1.0\textwidth]{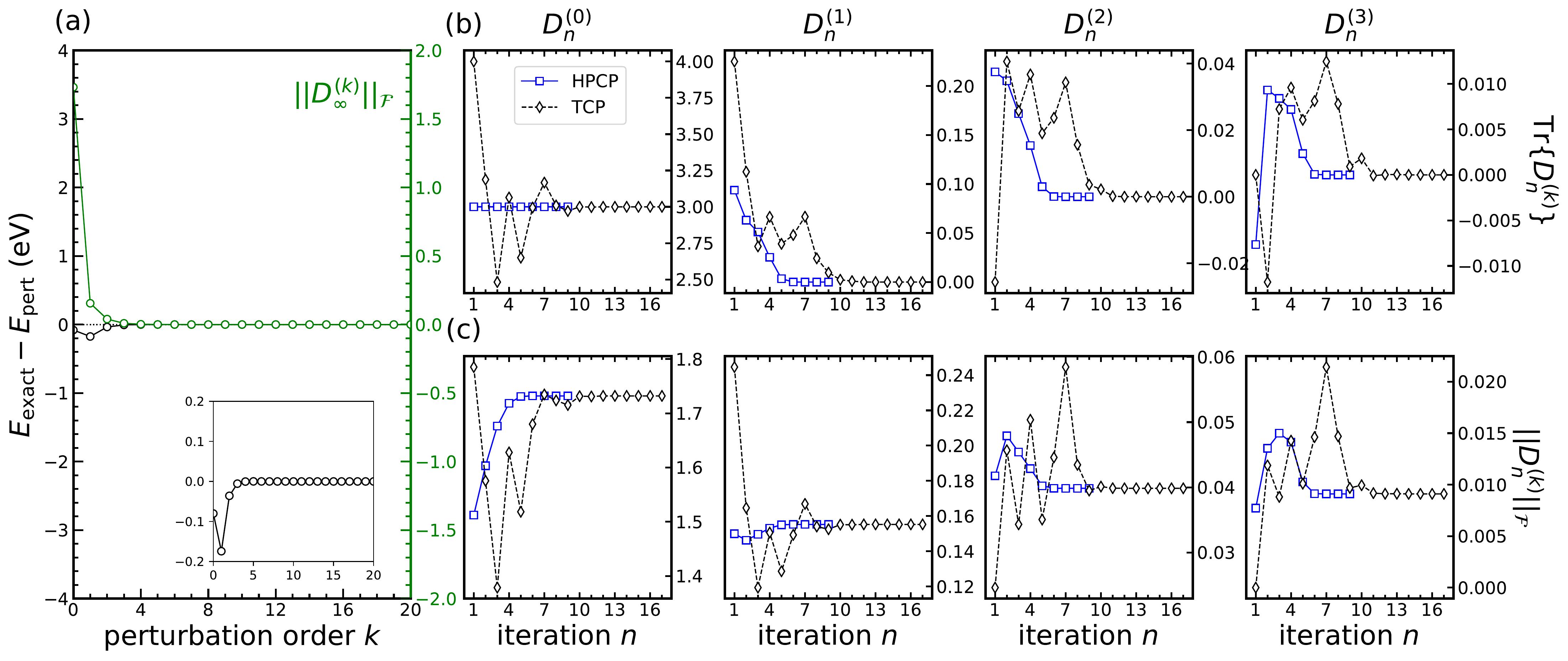} 
\caption{Perturbation series for $\pi$-bonding in pyridine (see text for details). (a) Convergence 
of the perturbed energy and Frobenius norm of the corresponding perturbed density 
matrices using purification-DMPT. The convergence of the unperturbed and firsts 3 perturbed 
density matrices during the purification cycles are presented in terms of their traces (b), and their 
norms (c), for both the TC2 and HPCP polynomials.}
\label{fig:pyridine}
\end{figure}
\end{widetext}

In both examples, the perturbation series is expected to converge such that, by setting $\lambda=1$, 
the sum over the perturbed densities in \eq{eq:Dmpert} converges towards the exact density matrix 
$D_{\lambda}$ as obtained from the full Hamiltonian $H_{\lambda}$ ; obtained for instance by 
diagonalizing $H_{\lambda}$ and summing over the $N$ occupied states as in \eq{eq:densop_idemp}. 
The same way, the sum over the perturbed energies, $E_{\trm{pert}} = \sum_k E^{(k)}$, computed 
from \eq{eq:fockener_pert} must converge towards the exact reference value, 
$E_{\trm{exact}} = 2\trace{H_{\lambda}D_{\lambda}}$. This is demonstrated for benzene and 
pyridine in  Figs.~\ref{fig:benzene}(a) and \ref{fig:pyridine}(a), respectively, where $E_{\trm{exact}}-E_{\trm{pert}}$ 
are plotted as a function of the number of perturbation terms entering in $\sum_k E^{(k)}$, up to $k=20$.
For benzene, we note that for symmetry reason only even orders contribute to $E_{\trm{pert}}$. 
At $k=20$, for benzene, we found that $E_{\trm{pert}}$ is converged to within 5 meV for 
$E_{\trm{exact}} = -88.9440$ eV, compared to a  much faster convergence for pyridine with 
the same level of convergence reach at $k=3$, with $E_{\trm{exact}} = -85.1005$ eV.
On the same figures ($y$-right axis) are also reported the norms of the density matrices 
$||D_\infty^{(k)}||$ as a function of $k$ which confirm the faster convergence of the
$\pi$-bonding perturbation in pyridine. 
We emphasize that identical results were obtained
with the reference McWeeny- and Sylvester-DMPT. Considering the benzene perturbation
series at higher order, reliability of the purification-based DMPT starts to degrade for $k>55$
using HPCP compared to $k>17$ for TC2. This trend has been confirmed for other systems
and $\pi$-bonding perturbations (not shown here), indicating that the HPCP purification
is more stable than TC2 when increasing the perturbation order. This could be related to the ill-definition
of the stopping criteria of TC2, becoming a very sensitive parameter in the extreme conditions 
of very high perturbation order. Nevertheless, this has a weak interest in practical applications
since they generally do not require a perturbed density matrix higher than the order 3.

Evolution of the density matrices (up to the order 3) during the purification process are plotted 
on Figs.~\ref{fig:benzene} and~\ref{fig:pyridine}, panels (b) for the trace and (c) for the norm.
First, in both cases, the convergence of the HPCP-DMPT is achieved in fewer iterations compared to 
TC2 as expected from the convergence profile of the two polynomials (quadratic vs. linear, respectively).
The property of trace-conservation of the zero-order density matrix fulfilled by the HPCP is apparent 
from these figures. In this respect, the TC2 polynomial demonstrates an erratic behavior with strong
oscillations at the beginning of the purification. These oscillations remain present at higher order, 
especially when looking to the pyridine case. We note that the zero-trace conservation of $D^{(1)}$ 
and $D^{(3)}$ observed with both polynomials for the case of benzene should not be overinterpreted 
since they are merely related to symmetry of the system. As a matter of fact, for the more general case 
of pyridine, such property of zero-trace conservation is not observed. Overall, the convergence 
behavior of the HPCP polynomials is more smooth than TC2  with, after the first few steps 
a quasi systematic monotonic approach of the converged perturbed density matrix.

\begin{figure}[h!]
\centering\includegraphics[width=0.45\textwidth]{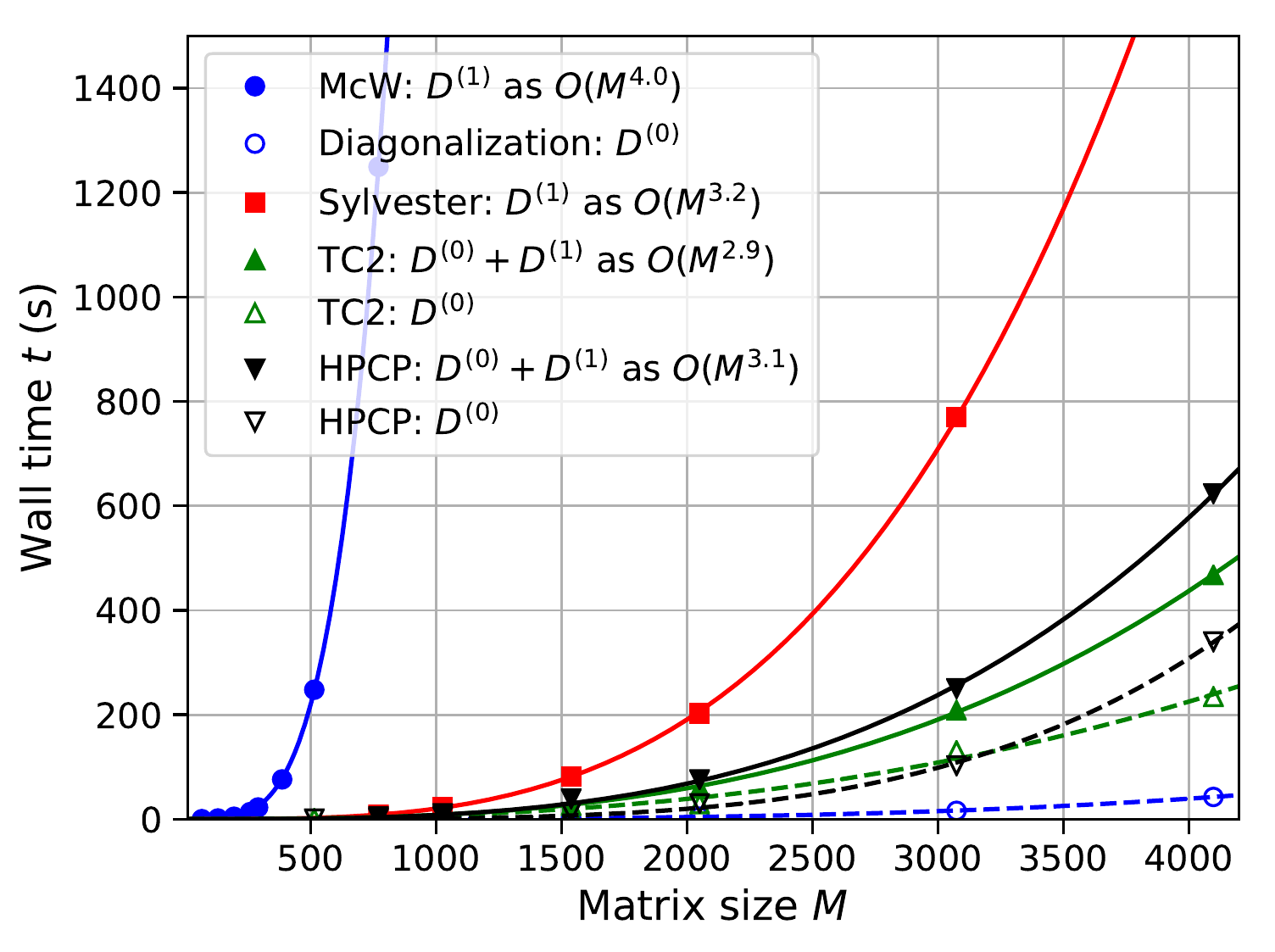} 
\caption{Scaling performance for the various methods to obtain $D^{(0)}$
and $D^{(1)}$ using a benchmark of H{\"u}ckel matrices of increasing size. 
Non-linear fits of equation $t = \alpha M^{\beta}$ are plotted with solid lines.}
\label{fig:scaling}
\end{figure}

In order to compare the computational performances of the purification,  McWeeny and Sylvester DMPT 
we have performed a set of calculations on systems of increasing size. As for the pyridine case, unperturbed 
H{\"u}ckel matrices for aromatic hydrocarbons with increasing ring size were generated along with first-order 
perturbed Hamiltonians describing the substitution of one of the carbon by a nitrogen atom. On Fig.~\ref{fig:scaling} 
is plotted the CPU time spent to obtain the unperturbed 
and first-order perturbed density matrices with respect to the size of the H{\"u}ckel matrix.
The same tolerence parameter of $||D_{n+1}-D_{n}||_{\mathcal{F}}<10^{-12}$ was used for purifications. 
We emphasize here that the McWeeny and Sylveter-DMPT are direct methods such that provided 
the eigenstates (or only the unperturbed density matrix for Sylveter-DMPT), the number of FLOPS to solve the DMPT 
equations is fixed by the size of the problem $M$, whereas for purification-based methods the solutions are found iteratively.
In this case the number of iterations depends on the band-gap of the system,\cite{Niklasson_TCP_2002,Rudberg_JPCM_2011}
and for HPCP, to a lesser extent, on the filling factor $\theta$.\cite{truflandier_jcp_2016} In our example, ideal 
conditions are fulfilled to minimize the number of purifications with $\theta=1/2$ and a large HOMO-LUMO gap 
when compared to the range of the full eigenspectrum. Note that the size of the N-substituted aromatic hydrocarbons 
does not impact the HOMO-LUMO gap, for which we found, as for pyridine, that the density matrices are converged after 
9 and 17 purifications for HPCP and TC2, respectively, independently of $M$. From Fig.~\ref{fig:scaling}, we note 
that exact diagonalization is more efficient than the purification methods to obtain the unperturbed density matrix. 
However, once we consider the perturbed state, we observe a net benefit to solve the Sylvester-DMPT equations 
using the Bartels-Stewart algorithm compare to the sum-over-states approach of McWeeny, which are expected to 
scales as $O(M^4)$ and $O(M^3)$, respectively. Note that the effective 
scaling reported on Fig.~\ref{fig:scaling} are in good agreement with these expectations. If now we consider purification
methods, which also scales as $O(M^3)$, the computational performances are further 
improved. We stress that for very low band gaps these performances are expected to degrade with a CPU
time multiplied by a factor 5 at worst.\cite{truflandier_jcp_2016} Finally, it should be mentioned that for 
purification-DMPT the plots of Fig.~\ref{fig:scaling} incorporate the calculation time for both $D^{(0)}$ 
and $D^{(1)}$ (since they cannot be decoupled), whereas for direct methods the former has not been included, 
increasing the interest for the TC2/HPCP DMPT. Comparison of the TC2 and HPCP shows that despite the more
rapid convergence of HPCP, the TC2 presents better performances due to its lower number of MMs
which is always (independently of the perturbation order) more than twice less than the one of HPCP. Typically 
in our example, TC2 requires 2 MMs per iteration compared to 5 for HPCP. However HPCP requires twice less iterations.

\section{Conclusions and perspectives}

In this work, we have reviewed three types of density matrix perturbation theory with, for
two of them, a resolution of the perturbed density matrices without the support of the unperturbed
eigenstates. These two methods, namely the Sylvester- and purification-DMPT,
clearly demonstrate better computational performances compared to the standard sum-over-states
approach. This indicates that current response equation solver as implemented in
quantum chemistry codes can be significantly accelerated using those two methods. We have
also sucessfully extended the recursive DMPT proposed by Niklasson and Challacombe to
the HPCP polynomial. Compared to TC2, the HPCP-DMPT shows a better stability when
approaching convergence, but remains more expensive in terms of computational performance.
We emphasize that, in favorable conditions, \ie\ for insulators with a filling factor
around 1/2, the TCP/HPCP-DMPT clearly outperforms the other approaches.
In the near future, we plan to implement the Sylvester- and purification-DMPT using non-orthogonal and
perturbation dependent basis set.\cite{niklasson_jcp_2005,niklasson_jcp_2007} Within the 
framework of linear scaling density functional theory as implemented in the CONQUEST code,\cite{conquest_2020} 
this will allow for application to electric and magnetic response calculations at a linear scaling computational cost.
\section{Data availability}
The data that support the findings of this study are available from the corresponding author upon reasonable request

\appendix
\section{SOS-McWeeny DMPT: 2nd and 3rd orders}
\label{app:sos-dmpt}

The second-order equation can be derived by applying the resolution of identity to
both side of \eq{eq:idemexpb}. Conserving notations of \eq{eq:mcwresolve}, we obtain
\begin{align}
 2\Dmook{2} +\Dmovk{2} +\Dmvok{2} &+  (\Dmk{1}\Dmk{1})_{oo} + (\Dmk{1}\Dmk{1})_{vv} \nonumber \\
   \cdots &+ (\Dmk{1}\Dmk{1})_{ov} + (\Dmk{1}\Dmk{1})_{vo} \nonumber  \\
   = \Dmook{2} &+ \Dmovk{2} + \Dmvok{2} + \Dmvvk{2} 
 \label{second_one}          
\end{align}
By resolving the product of first-order perturbed density matrices according to
\begin{align}
   \Dmk{1}\Dmk{1}  & =  \Dmk{1} I \Dmk{1} \nonumber\\
                              & =   \Dmk{1} (\Dm + \Db) \Dmk{1}\nonumber\\
                              & =   \Dmk{1} (\Dm^2 + \Db^2) \Dmk{1}\nonumber
\end{align}
we obtain
\begin{subequations}
\label{sec_xx} 
\begin{align}
(\Dmk{1}\Dmk{1})_{oo} & = \Dmook{1}\Dmook{1} + \Dmovk{1}\Dmvok{1} \label{sec_oo} \\
(\Dmk{1}\Dmk{1})_{vv} & = \Dmvvk{1}\Dmvvk{1}  + \Dmvok{1}\Dmovk{1} \label{sec_vv} \\
(\Dmk{1}\Dmk{1})_{ov} & = \Dmovk{1}\Dmvvk{1} + \Dmook{1}\Dmovk{1} \label{sec_ov} \\
(\Dmk{1}\Dmk{1})_{vo} & = \Dmvok{1}\Dmook{1} + \Dmvvk{1}\Dmvok{1} \label{sec_vo}          
\end{align}
\end{subequations}
On inserting the rhs of \eqs{sec_xx} into \eqref{second_one}, 
and using the properties of \eq{eq:mcwresolvee}, we have
\begin{align}
	2\Dmook{2}+\Dmovk{2}&+\Dmvok{2}+\Dmovk{1}\Dmvok{1}+\Dmvok{1}\Dmovk{1} \nonumber\\
	&= \Dmook{2} + \Dmovk{2} + \Dmvok{2} + \Dmvvk{2}
\end{align}
Therefore, we find
\begin{subequations}
\label{sec_offdiag}
\begin{align}
	\Dmook{2} &= -\Dmovk{1}\Dmvok{1} \\
	\Dmvvk{2} &= +\Dmvok{1}\Dmovk{1} 
\end{align}
\end{subequations}
Unlike the 1st-order perturbation, the diagonal components of the 2nd-order perturbed density matrix are
likely to be non-zero and can be computed from the 1st-order perturbed density matrix. Relying 
furthermore on the symmetry of the perturbed density, it leaves only the occupied-virtual 
coupling block matrix to evaluate. On resolving the 2nd-order perturbed Hamiltonian matrix using 
\eq{eq:scfexpb}, we obtain
\begin{equation}
\Fmovk{2} = [\Fm,\Dmovk{2}] + [\Fmk{1},\Dmk{1}]_{ov} \label{sec_scf1}	
\end{equation}
Using the spectral resolution of the non-perturbed Hamiltonian matrix and the perturbed density matrix, 
\eq{sec_scf1} transforms as
\begin{equation}
\Fmovk{2} = \sum_{i,j}  \left( \Dmovijk{2} (\Em_i - \Eb_j) 
+ [\Fmk{1},\Dmk{1}]_{ov,ij} \right)
\end{equation}
which leads to
\begin{equation}
\Dmovk{2} = \sum_{i,j} (\Em_i - \Eb_j )^{-1} \left( \Fmovk{2} - [\Fmk{1},\Dmk{1}]_{ov} \right)_{ij} \label{sec_dm2}
\end{equation}
The final 2nd-order perturbed density matrix is obtained by summing over $\Dmovk{2}$, 
its conjugate-transposed and the block-diagonal contributions of \eq{sec_offdiag}.

Using the same route, the third-order response equation can be derived from \eq{eq:idemexpc}.
This yields to
\begin{align}
 2\Dmook{3}+\Dmovk{3}+\Dmvok{3} &+  (\Dmk{1}\Dmk{2})_{oo}+(\Dmk{1}\Dmk{2})_{vv} \nonumber \\
   \cdots &+ (\Dmk{1}\Dmk{2})_{ov}+ (\Dmk{1}\Dmk{2})_{vo}  \nonumber  \\
   \cdots &+ (\Dmk{2}\Dmk{1})_{oo}+(\Dmk{2}\Dmk{1})_{vv}   \nonumber  \\
   \cdots &+ (\Dmk{2}\Dmk{1})_{ov}+(\Dmk{2}\Dmk{1})_{vo}   \nonumber  \\
   = \Dmook{3} &+ \Dmovk{3} + \Dmvok{3} + \Dmvvk{3} \label{third_one}        
\end{align}
where
\begin{subequations}
\label{third_sms}
\begin{align}
(\Dmk{1}\Dmk{2})_{oo} = \Dmook{1}\Dmook{2} + \Dmovk{1}\Dmvok{2} \label{thd_oo} \\
(\Dmk{1}\Dmk{2})_{vv} = \Dmvvk{1}\Dmvvk{2} + \Dmvok{1}\Dmovk{2} \label{thd_vv} \\
(\Dmk{1}\Dmk{2})_{ov} = \Dmovk{1}\Dmvvk{2} + \Dmook{1}\Dmovk{2} \label{thd_ov} \\
(\Dmk{1}\Dmk{2})_{vo} = \Dmvok{1}\Dmook{2} + \Dmvvk{1}\Dmvok{2} \label{thd_vo} \\
(\Dmk{2}\Dmk{1})_{oo} = \Dmook{2}\Dmook{1} + \Dmovk{2}\Dmvok{1} \label{thd_oop} \\
(\Dmk{2}\Dmk{1})_{vv} = \Dmvvk{2}\Dmvvk{1} + \Dmvok{2}\Dmovk{1} \label{thd_vvp} \\
(\Dmk{2}\Dmk{1})_{ov} = \Dmovk{2}\Dmvvk{1} + \Dmook{2}\Dmovk{1} \label{thd_ovp} \\
(\Dmk{2}\Dmk{1})_{vo} = \Dmvok{2}\Dmook{1} + \Dmvvk{2}\Dmvok{1} \label{thd_vop}           
\end{align}
\end{subequations}
From \eqs{eq:mcwresolvee}, \eqref{sec_offdiag} and \eqref{third_sms},
\eq{third_one} simplifies to
\begin{align}
2\Dmook{3} + \Dmovk{3} + \Dmvok{3} &+ \Dmovk{1}\Dmvok{2} + \Dmovk{2}\Dmvok{1} \nonumber\\
\cdots &+ \Dmvok{1}\Dmovk{2}+\Dmvok{2}\Dmovk{1} \nonumber\\
= \Dmook{3} &+ \Dmovk{3} + \Dmvok{3} + \Dmvvk{3} \label{thrd_one}
\end{align}
On identifying lhs and rhs terms, it follows that
\begin{subequations}
\label{thrd_offdiag}
\begin{align}
	\Dmook{3} &=  -\left(\Dmovk{1}\Dmvok{2}+\Dmovk{2}\Dmvok{1}\right) \\
	\Dmvvk{3} &=  +\left(\Dmvok{1}\Dmovk{2}+\Dmvok{2}\Dmovk{1}\right) 
\end{align}
\end{subequations}
Again, the last equation shows that the diagonal components are computed with the
1st and 2nd-order perturbed density matrices. At this point, we emphasize that only the occupied-virtual 
transition matrix needs to be evaluated since the perturbed density matrix 
is Hermitian. 

Relying on the spectral resolution, the 3rd-order perturbed Hamiltonian matrix is given by
\begin{subequations}
\begin{align}
\Fmovk{3} &= [\Fm,\Dmovk{3}] + [\Fmk{1},\Dmk{2}]_{ov} + [\Fmk{2}\Dmk{1}]_{ov}\label{thrd_scf1} \\
& = \sum_{i,j} \Dmovijk{3} (\Em_i - \Eb_j) \nonumber\\
&\times\left(  [\Fmk{1},\Dmk{2}]_{ov} + [\Fmk{2}\Dmk{1}]_{ov} \right)_{ij}
\end{align}
\end{subequations}
which leads to
\begin{align}
\Dmovk{3} &= \sum_{i,j} (\Em_i - \Eb_j )^{-1}\nonumber\\
                 &\times\left( \Fmovk{3} - [\Fmk{1},\Dmk{2}]_{ov}
                 - [\Fmk{2},\Dmk{1}]_{ov} \right)_{ij}\label{thrd_dm3}
\end{align}
By summing over contributions of \eqs{thrd_offdiag} and \reff{thrd_dm3} and its conjugate-transposed,
we obtain the 3rd-order perturbed density matrix. It is worth mentioning that the direct resolutions of the 2nd 
or 3rd-order perturbed quantities implies prior knowledge of the lower orders (1st and 2nd, respectively), in such
a way that, whatever is the order to be resolved, \ie\ \eqs{thrd_offdiag}--\reff{thrd_dm3} on one side, or 
\eqs{sec_offdiag}--\reff{sec_dm2} on the other side, both cases involve to solve for the linear-response 
of the order $(k-1)$ to obtain the perturbed quantities at the order $k$.
Eventually, by mathematical induction it is straightforward to generalize these working equations to any
order $k$, as introduced in \eqs{dbov1_dcpscf}--\reff{dbvv_dcpscf}.

\section{Alternative formulation of DMPT}
\label{app:lm-dmpt}
In this appendix, we shall show that the planewave-based DMPT introduced 
by Lazzeri and Mauri\cite{lazzeri_high-order_2003} is closely related to the 
atomic-orbital based DMPT method of Kussmann and Ochsenfeld\cite{kussmann_jcp_2007-1,kussmann_jcp_2007-2},
(and by extension the Sylvester-DMPT) both relying on the CG minimization. Thereafter, 
they will be referred to as PW-DMPT and AO-DMPT, respectively.

As discussed in \sec{sec:cg-dcpscf}, employing a PW basis set with
$M_{\textrm{PW}}>>N$ constrains us to use iterative diagonalization,
where, for an insulator, the $N$ first (lowest) eigenstates, $\{\epsilon_i,\psi_i\}^N_{i\in\trm{occ}}$,
necessary and sufficient to obtain the unperturbed ground-state, can be determined 
with a satisfying accuracy. Here, we will assume the linear-DMPT regime: in order to 
determine the $k$th-order perturbed density, all the preceding orders, up to $(k-1)$, 
are known. As a result, for a PW basis set, at the zero order, the unperturbed density matrix, 
$\Dmk{0}\equiv\Dm$, can be resolved in terms of the occupied states following \eq{eq:densop_idemp},
and knowing only the occupied eigenstates, the unperturbed Hamiltonian matrix, $\Fmk{0}\equiv\Fm$, 
can be formally expressed as:
\begin{equation}
   \Fm = \sum_{i\in\trm{occ}} |\psi_i\rangle\epsilon_i\langle\psi_i| + \Db\Fm\Db 
   \label{eq:lm_app0}
\end{equation}
recalling that, from the closure relation of \eq{eq:closure}, $\Db = I - \Dm$.
Within the framework of the AO-DMPT method, the $k$-order perturbed 
density matrix, $\Dmk{k}$, is found as solution of the following equation:
\begin{equation}
	[\Dm, \sum^{k}_{l=0}[\Fmk{l},\Dmk{k-l}]] = 0 
\label{eq:lm_app1}
\end{equation}
It must be emphasized that, in comparison to the McWeeny-DMPT
and \eq{eq:mcweenyfocka}, or the PW-DMPT and \eq{eq:lm_app0},
the resolution of the AO-DMPT equation is free of any intermediate spectral 
resolution of the unperturbed Hamiltonian matrix.
Indeed, the PW-DMPT may be view as an intermediate
strategy, between the former and the later, where \eq{eq:lm_app1} 
is decomposed to perform an occupied-perturbed state-by-state 
resolution.\cite{lazzeri_high-order_2003} For instance, on applying 
to \eq{eq:lm_app1} the identity: $[\Dm,O] = \Dm O\Db - \Db O\Dm$, 
and projecting to the right on $|\psi_i\rangle$ with $i\in\trm{occ}$, 
we find:
\begin{equation}
	\Db \sum^{k}_{l=0}[\Fmk{l},\Dmk{k-l}] |\psi_i\rangle  = 0
\label{eq:lm_app2}	
\end{equation}
By extracting the terms containing the $k$th-order density matrix from the sum, 
and by reordering, it appears:
\begin{equation}
	\left( \Db\Fm\Dmk{k} - \Db\Dmk{k}\Fm\right)|\psi_i\rangle 
	= -\sum^{k}_{l=1}\Db[\Fmk{l},\Dmk{k-l}]|\psi_i\rangle
\label{eq:lm_app3}	
\end{equation}
On substitution of the expression of $\Fm$ from \eq{eq:lm_app0} into
\eq{eq:lm_app3}, and recalling that, at the zero temperature limit, 
the converged ground-state one-electron and one-hole density matrices
must respect the idempotency and the stationary conditions of \eqs{eq:idemexp0} 
and \reff{eq:scfexp0}, respectively, we arrive at:
\begin{equation}
	\left( \Fm\Db\Dmk{k} - \Db\Dmk{k}\epsilon_i\right)|\psi_i\rangle 
	= -\sum^{k}_{l=1}\Db[\Fmk{l},\Dmk{k-l}]|\psi_i\rangle
\label{eq:lm_app4}	
\end{equation}
Complying with notations of \citen{lazzeri_high-order_2003} by introducing,
\begin{subequations}
\label{eq:lm_app5}
 \begin{align}
	|\eta^{(k)}_{i}\rangle &= \Db\Dmk{k}|\psi_i\rangle \label{eq:lm_app5a}\\
    \trm{such that}\quad\Dmvok{k} &= \sum_{i\in\trm{occ}}|\eta^{(k)}_{i}\rangle\langle\psi_i |\label{eq:lm_app5b}
 \end{align}
\end{subequations}
we recover Eq. (13) of the article, that is,
\begin{equation}
	\left( \Fm - \Id\epsilon_i\right)|\eta^{(k)}_i\rangle 
	= -\sum^{k}_{l=1}\Db[\Fmk{l},\Dmk{k-l}]|\psi_i\rangle
\label{eq:lm_app6}	
\end{equation}
In the interests of completeness, the SOS equation \reff{dbov3_dcpscf} can be 
found back by further resolving the full spectrum of $\Fm$. By using \eq{eq:mcweenyfocka},
along with the resolution of identity \reff{eq:closure}, the lhs of \eq{eq:lm_app6} 
transforms as
\begin{align}
	\left( \Fm - \epsilon_{i}\Id\right)|\eta^{(k)}_{i}\rangle 
	&= \sum_{i'\in\trm{occ}}   |\psi_{i'}\rangle(\epsilon_{i'} - \epsilon_{i})\langle\psi_{i'} |\eta^{(k)}_{i}\rangle  \nonumber\\
	&+ \sum_{j'\in\trm{virt}} |\bar{\psi}_{j'} \rangle(\bar{\epsilon}_{j'} - \epsilon_{i})\langle\bar{\psi}_{j'} |\eta^{(k)}_{i} \rangle
\label{eq:lm_app7}	
\end{align}
By remarking that: (\textit{i}) $\langle\psi_{i'} |\eta^{(k)}_{i}\rangle=0$ $\forall i'$,
and (\textit{ii})  $\langle\bar{\psi}_{j'}|\eta^{(k)}_{i}\rangle=\langle\bar{\psi}_{j'}|\Dmk{k}|\psi_{i}\rangle$,
we found that \eq{eq:lm_app7} simplifies to
\begin{equation}
	\left( \Fm - \epsilon_{i}\Id\right)|\eta^{(k)}_{i}\rangle = 
	\sum_{j\in\trm{virt}} |\bar{\psi}_{j} 
	\rangle(\bar{\epsilon}_{j} - \epsilon_{i})
	\langle\bar{\psi}_{j} | \Dmk{k} | \psi_{i}\rangle
\label{eq:lm_app8}	
\end{equation}
Inserted in the lhs of \eq{eq:lm_app6} and resolving the one-hole density matrix of the rhs,
we have
\begin{align}
	\sum_{j\in\trm{virt}} &(\bar{\epsilon}_{j} - \epsilon_{i})
	|\bar{\psi}_{j}\rangle\langle\bar{\psi}_{j} | \Dmk{k} | \psi_{i}\rangle\nonumber\\
	&= \sum^{k}_{l=1}\sum_{j\in\trm{virt}} | \bar{\psi}_{j}\rangle\langle\bar{\psi}_{j} | 
	    [\Dmk{k-l},\Fmk{l}]|\psi_i\rangle
\label{eq:lm_app9}	
\end{align}
which further gives the analytical expression of $|\eta^{(k)}_{i}\rangle$ 
with respect to the lower order perturbed density matrices,
\begin{equation}
	|\eta^{(k)}_{i}\rangle
	= \sum^{k}_{l=1}\sum_{j\in\trm{virt}}\frac{\langle\bar{\psi}_{j} | 
	    [\Dmk{k-l},\Fmk{l}]|\psi_i\rangle}{\bar{\epsilon}_{j}-\epsilon_{i}} | \bar{\psi}_{j}\rangle
\label{eq:lm_app10}	
\end{equation}
Following definitions \reff{eq:lm_app5}, by summing over the $N$ perturbed projectors, 
this yields to the $k$th virtual-occupied transition matrix,
\begin{equation}
	\Dmvok{k} = \sum^{k}_{l=1}\sum_{i\in\trm{occ}}\sum_{j\in\trm{virt}}\frac{\langle\bar{\psi}_{j} | 
	    [\Dmk{k-l},\Fmk{l}]|\psi_i\rangle}{\bar{\epsilon}_{j}-\epsilon_{i}} | \bar{\psi}_{j}\rangle\langle\psi_i|
\label{eq:lm_app10}	
\end{equation}
which is the conjugate transpose of the McWeeny equation \reff{dbov3_dcpscf}. 
Note that, for $k>1$, the remaining occupied-occupied and virtual-virtual 
components necessary to build the full $k$th-order matrix, \ie\ $\Dmk{k} 
= \Dmovk{k} + \Dmvok{k} + \Dmook{k} + \Dmvvk{k}$, can be easily 
computed from the lowest orders using \eqs{dboo_dcpscf} and \reff{dbvv_dcpscf}.
%
%
%
%
%
%
\end{document}